\documentclass[11pt]{article}
\usepackage[margin=1in]{geometry}
\usepackage{amsmath,amssymb,amsthm}
\usepackage{graphicx}
\usepackage{array}
\usepackage{enumitem}
\usepackage{empheq}
\usepackage{color}
\usepackage{cite}
\usepackage{hyperref}
\usepackage{authblk}
\usepackage[section]{placeins}
\usepackage{tikz}
\usetikzlibrary{positioning,calc,arrows.meta,fit}
\newcommand{\doubleangle}[1]{\langle\kern-0.2em\langle #1 \rangle\kern-0.2em\rangle}

\newcommand{\pv}{\mbox{;}\,}
\providecommand{\lblot}{\langle\!|}
\providecommand{\rblot}{|\!\rangle}
\newcommand{\ccdot}{\hspace{-4pt}\cdot\hspace{-4pt}}
\newcommand{\conc}{\rblot\ccdot\lblot}

\makeatletter\@ifundefined{theorem}{\newtheorem{theorem}{Theorem}}{}\makeatother
\newtheorem{proposition}{Proposition}
\newtheorem{corollary}{Corollary}
\newtheorem{lemma}{Lemma}
\newtheorem{remark}{Remark}
\newtheorem{result}{Result}
\newenvironment{summary}%
  {\par\medskip\noindent\textit{Summary.}\ \ignorespaces}%
  {\par\smallskip\centerline{\rule{0.3\linewidth}{0.4pt}}\par\medskip}
\newenvironment{inwords}%
  {\par\smallskip\noindent\textit{Physical content.}\ \ignorespaces}%
  {\par\smallskip}

\begin{document}

\title{Memory kernel and closure for systems driven by intermittent forcing
with power-law waiting times}

\author[1]{M.~Bianucci\thanks{Corresponding author. E-mail: marco.bianucci@cnr.it}}
\affil[1]{Istituto di Scienze Marine, Consiglio Nazionale delle Ricerche (ISMAR--CNR), 19032 Lerici (SP), Italy}
\author[2]{R.~Mannella}
\affil[2]{Dipartimento di Fisica, Universit\`a di Pisa, 56100 Pisa, Italy}

\date{\today}
\maketitle

\begin{abstract}
Many systems are driven by an intermittent forcing that takes on a random value for a
random time interval. For such renewal step noise, the master equation for the driven
variable has a memory kernel whose coefficients are the totally time-ordered cumulants,
or $G$-cumulants, of the noise. In the case of Poisson switching, these correspond to the
Boolean cumulants of the jump distribution, and the kernel can be summed exactly. We
derive an exact decomposition of these coefficients for any waiting-time distribution
with independent amplitudes and durations, which separates the jump statistics from the
renewal statistics. For power-law waiting-time densities,
$\psi(\theta)\propto\theta^{-\mu}$ with $\mu>2$ (finite mean), a long time interval is
most likely covered by a single epoch, and we prove that, as a result, the coefficients
cross over at long lags, at all orders, from the Boolean cumulants to the ordinary
moments of the jump law. The long-lag kernel thus becomes a superposition of the kernels
of the frozen deterministic flows, weighted by $\xi^2p(\xi)$, and the corresponding
master equation follows in closed form. For $2<\mu<3$, where the noise correlation time
is infinite, the second-order closure retains a finite error even for slow dynamics,
whereas the frozen-flow master equation yields the exact stationary mean in that limit.
The telegraph process is the only renewal step noise for which the second-order closure
is exact. An exactly solvable model with correlated additive and multiplicative noise
confirms these predictions.
\end{abstract}


\tableofcontents

\section{Introduction\label{sec:intro}}

Many systems in physics, the geosciences, biology, and the social sciences are governed by
an intermittent driving force. Rather than fluctuating continuously, this driver remains
approximately constant for a random time interval before jumping abruptly to a new value.
Examples of this phenomenon include the stochastic velocity of a Lévy
walk~\cite{swkPRL58,zdkRMP87}, the on/off states of a nanoscale blinking
emitter~\cite{kfhgJCP112,shbPT62}, the laminar and turbulent phases of an intermittent
flow~\cite{pmCMP74,mJP41}, and episodic atmospheric forcing, such as bursts of westerly
winds that influence the El Niño--Southern Oscillation (ENSO)~\cite{eytJC18,jltzGRL34}.

A minimal representation of such a forcing is a renewal process $\xi[t]$. At each renewal,
an amplitude $\xi$ is chosen according to a density function $p(\xi)$ and is maintained
for a waiting time $\theta$, which is in turn drawn from a density function
$\psi(\theta)$; the amplitude and the duration are independent of each other, and
successive pairs of amplitude and duration are independent and identically
distributed~\cite{zdkRMP87,bblmCSF196}. The law governing the jumps determines the value
taken by the forcing, while the law governing the waiting times establishes the duration
of each of these values. A variable $x$, driven by a forcing of this type, follows
\begin{equation}
\dot x=-C(x)+\epsilon I(x)\xi[t],
\label{eq:SDE}
\end{equation}
where $-C(x)$ is the unperturbed velocity field, while $I(x)$ describes the state-dependence of the noise intensity.
The central problem lies in determining which properties of $p(\xi)$ and $\psi(\theta)$ are preserved
in the reduced dynamics of $x$, that is, in an evolution equation for the probability density function (PDF)
of the variable $x$ alone, which retains the multi-time statistics of the driving signal without requiring
the explicit tracking of its individual realizations.

For a given realization of the forcing, the density $\rho(x\pv t)$ of an
ensemble of trajectories obeys the stochastic Liouville equation
\begin{equation}
\partial_t\rho(x\pv t)
 =
\big[\mathcal L_a+\epsilon\,\xi[t]\,\mathcal L_I\big]\rho(x\pv t),
\qquad
\mathcal L_a\rho:=\partial_x\big[C(x)\rho\big],\quad
\mathcal L_I\rho:=-\partial_x\big[I(x)\rho\big].
\label{eq:liouville}
\end{equation}
Averaging over the realizations of $\xi[t]$, by projection-operator
methods~\cite{Zwanzig2001,grigolini1989,tPhysica74,chkPNAS} or generalized
cumulant expansions~\cite{kuboGenCumJPSJ17,kuboGenCumJMP4,fJCP49,ydfJCP62,
vkPhysica74a,vkPhysica74b,fJMP17,fJMP20,sthJSP17} (see
Refs.~\cite{fFPReport1978,hjACP89,vanKampen2007} for reviews), leads to an
exact evolution equation for the  reduced PDF
$P(x\pv t):=\langle\rho(x\pv t)\rangle$ containing a memory kernel. For an
initial density independent of the forcing,
\begin{equation}
\partial_tP(x\pv t)
 =
\mathcal L_aP(x\pv t)
 +
\int_{t_0}^t\mathcal G(t,u)\,P(x\pv u)\,\mathrm du ,
\label{eq:ME}
\end{equation}
where $t_0$ is the initial time. The kernel can be expanded in powers of
$\epsilon$, with coefficients given by the totally time-ordered, or
$G$-, cumulants of the forcing. These cumulants are therefore the channel
through which the statistics of the renewal process enter the reduced
dynamics. Their precise definition and the corresponding all-order kernel
expansion are given in Section~\ref{sec:setting}.

In conventional perturbation-theoretic approaches, the memory term in Eq.~\eqref{eq:ME} is
eliminated by expressing the density at an earlier time via backward evolution and
expanding the result in powers of $\epsilon$. This yields a temporally local equation,
which is typically truncated at second order~\cite{sthJSP17}. Such a truncation is exact
in the case of Gaussian white noise; for colored or non-Gaussian noise, it is often
justified when the dynamics under consideration are slow compared to the correlation time
of the noise. However, for renewal processes with waiting times that follow a power
law $\psi(\theta)\sim\theta^{-\mu}$, the correlation time of the noise is infinite when
$2<\mu<3$. Furthermore, even in cases where this time is finite, longer time intervals may
still remain significant. Consequently, an argument based on the separation of time scales
is insufficient to guarantee the accuracy of a low-order closure.

For certain Markov processes, specifically, dichotomous Markov
noise~\cite{slPhysicaA91,vdbJSP31,Horsthemke1984, bIJMPB20} and kangaroo
processes~\cite{BrissaudFrisch1974}, exact higher-order results are available. In the case
of non-exponential renewal processes, however, the driving force is non-Markovian; its
higher-order correlations generally cannot be factorized and, prior to reaching the steady
state, depend on the age of the process~\cite{agpwPRE70,aagprPRE68}. For non-exponential
renewal processes, researchers have investigated generalized master equations, linear
response theory, and stochastic
resonance~\cite{aagprPRE68,ghPRL91,ghPRE69,sPRE73,fPRE50,fPRE52}; furthermore, within the
framework of the Kubo--Anderson line-shape theory, studies have examined residence times
characterized by a power-law distribution~\cite{jbsCP284,apgPRL93}. Analogous kinetic
descriptions are found in continuous-time random walks
(CTRWs)~\cite{mwJMP6,smPRB12,mkPR339} and Lévy walks~\cite{zdkRMP87}. What has been
missing thus far is a general theoretical framework that, at all orders, separates jump
statistics from renewal statistics by operating at the level of the memory
kernel---specifically, through what are known as $G$-cumulants. Such a theoretical framework
would make it possible to evaluate the accuracy of a lower-order approximation, rather
than simply taking it for granted.

The theoretical framework developed here is based on the formalism of generalized
cumulants (or $M$-cumulants), which was introduced in the works
\cite{bbJSTAT4,bCSF148,bCSF159}. In this context, total time ordering defines the
$G$-cumulants that appear in the memory kernel, whereas partial time ordering leads to a
temporally local equation. Renewal processes provide a natural application for this: in
\cite{bbmJSTAT2026}, exact and asymptotic results were obtained for a continuous-time
random walk with drift and position-dependent noise intensity; meanwhile, in
\cite{bblmCSF196,bblmCSF202}, universal correlation structures for two or more time points
were derived. These include a representation of the correlation functions as a sum over
interval partitions of the observation times, as well as a reduction to the extreme times
in cases where waiting times follow a power law and possess a finite mean.

The immediate starting point for the present work is the exact Boolean structure derived
in \cite{bmBoolean} for exponential waiting times. In this case, the renewal process is
Markovian, and renewal events occurring in disjoint time intervals are statistically
independent. The $G$-cumulants correspond to the Boolean cumulants of the jump
distribution, multiplied by the probability that all observation time points fall within a
single renewal interval; this allows for an exact summation of the memory kernel. The
result consists of an average of the kernels corresponding to so-called \emph{frozen
flows}---that is, deterministic flows calculated for fixed values of $\xi$---where the
averaging is performed with respect to the Boolean Lévy measure associated with the jump
law; this measure does not coincide with the distribution of the values taken on by the
noise.

Power-law waiting times between successive switches of a forcing are observed in a wide
range of contexts: in intermittent dynamics~\cite{pmCMP74,mJP41}, in
Lévy walks and anomalous transport~\cite{swkPRL58,zkPRE47,swsPRL71,zdkRMP87}, in
nanostructures characterized by intermittent emission (so-called ``blinking
nanostructures'')~\cite{kfhgJCP112,bhmdPRL90,shbPT62}, in sub-recoil laser
cooling~\cite{bbeaPRL72}, and even in precipitation events~\cite{phcPRL88}. Furthermore,
they are widespread in datasets from biological, social, and financial domains
(see~\cite{zdkRMP87,wg2021} and the literature cited therein). For $1<\mu<2$, the mean
waiting time diverges; in such cases, the process exhibits aging effects and weak
ergodicity breaking~\cite{bJPI2,bPRL90,bbPRL94}. We examine the case where $\mu>2$ (or
$\mu=2$ with a finite mean value), for which a stationary regime exists.

From the perspective of probability theory, power laws are
subexponential~\cite{foss2013,embrechts1997}: an unusually large sum of waiting times is
typically driven by a single, exceptionally large contribution. In physics, the
single-big-jump principle has emerged as a modeling tool for rare events---for instance,
in Lévy walks, in cold atoms within optical lattices, in the L\'evy--Lorentz gas, and in
correlated random walks~\cite{vbbPRE100,hbEPJB94}, as well as in transport within
disordered media, where the longest trapping time determines rare fluctuations and even
the mean squared displacement~\cite{wvbbPRR1}. Such individual events give rise to
infinite densities, which define the higher-order moments of Lévy walks~\cite{rdhbPRE90}.
These studies focus on the far tails of additive observables, such as
displacements or sums. In the case under consideration, the principle applies to the
durations of the epochs of a forcing that drives a dissipative system; it
thus enters directly into the exact reduced dynamics.

The unifying quantity is
\begin{equation}
P_1(t_i,t_j)
 =
\Pr\!\left[\text{no renewal occurs in }[t_i,t_j]\right],
\label{eq:P1_intro}
\end{equation}
the probability that a single renewal epoch covers the observation interval.
For stationary renewal with finite mean waiting time it depends only on the
span $\Delta=t_j-t_i$, and we write $P_1(\Delta)$. For power-law waiting times
the long-lag properties of the reduced dynamics are dominated by histories in
which one exceptionally long epoch covers the memory interval. We refer to
this as the \emph{single-epoch mechanism}, and show the following.
\begin{enumerate}[label=(\roman*)]
\item For any waiting-time law the $G$-cumulants separate exactly into
Boolean cumulants of the jump law and weights that depend only on the renewal
process. At second order the kernel depends on the waiting times only through
$P_1$, and the telegraph process is the only renewal step noise for which a
second-order closure is exact (Section~\ref{sec:decomposition}).
\item For power-law waiting times with finite mean the coefficients cross
over at long lags, at all orders, from the Boolean cumulants to the ordinary
moments of the jump law (Section~\ref{sec:powerlaw}).
\item As a consequence, the long-lag kernel becomes a superposition of
frozen-flow kernels weighted by $\xi^2p(\xi)$, and the master equation takes a
closed form, Eq.~\eqref{eq:ME_mom} (Section~\ref{sec:ht_kernel}).
\item In an exactly solvable model, for $2<\mu<3$ a second-order closure keeps
a finite error however slow the dynamics, whereas the frozen-flow master
equation becomes exact in that limit (Section~\ref{sec:limi}).
\end{enumerate}

\begin{figure}[tp]
\centering
\resizebox{0.80\textwidth}{!}{%
\begin{tikzpicture}[
    node distance=0.4cm and 0.55cm,
    every node/.style={font=\footnotesize,
        execute at begin node={\hyphenpenalty=10000}},
    mainbox/.style={
        draw,
        rounded corners=2pt,
        fill=gray!14,
        align=center,
        text width=6.2cm,
        minimum height=0.55cm,
        inner sep=5pt
    },
    inputbox/.style={
        draw,
        rounded corners=2pt,
        fill=green!7,
        align=center,
        text width=3.5cm,
        minimum height=0.78cm,
        inner sep=4pt
    },
    routebox/.style={
        draw,
        rounded corners=2pt,
        fill=blue!7,
        align=center,
        text width=4.5cm,
        minimum height=0.9cm,
        inner sep=5pt
    },
    resultbox/.style={
        draw,
        rounded corners=2pt,
        fill=orange!8,
        align=center,
        text width=4.5cm,
        minimum height=0.92cm,
        inner sep=5pt
    },
    mergebox/.style={
        draw,
        very thick,
        rounded corners=2pt,
        fill=violet!8,
        align=center,
        text width=8.5cm,
        minimum height=0.95cm,
        inner sep=6pt
    },
    comparisonbox/.style={
        draw,
        dashed,
        rounded corners=2pt,
        fill=red!3,
        align=center,
        text width=4.3cm,
        minimum height=1.0cm,
        inner sep=5pt
    },
    extensionbox/.style={
        draw,
        dashed,
        rounded corners=2pt,
        fill=yellow!10,
        align=center,
        text width=4.15cm,
        minimum height=1.15cm,
        inner sep=5pt
    },
    arr/.style={-{Latex[length=2mm]},thick},
    dashedarr/.style={
        -{Latex[length=2mm]},
        thick,
        dashed
    }
]


\node[mainbox] (jointlaw)
{JOINT AMPLITUDE--DURATION LAW\\[1mm]
$q(\xi,\theta)$};

\node[extensionbox, right=0.85cm of jointlaw] (coupled)
{CORRELATED EXTENSION\\[1mm]
$q(\xi,\theta)\neq p(\xi)\psi(\theta)$\\
long epochs select $p_\Delta(\xi)$};

\draw[dashedarr] (jointlaw) -- (coupled);


\node[inputbox, below left=0.62cm and 0.22cm of jointlaw] (waiting)
{WAITING-TIME LAW\\
$\psi(\theta)$};

\node[inputbox, below right=0.62cm and 0.22cm of jointlaw] (jump)
{JUMP LAW\\
$p(\xi)$};

\node[font=\scriptsize, align=center]
at ($(waiting)!0.5!(jump)+(0,0.22)$)
{main analysis:
$q(\xi,\theta)=p(\xi)\psi(\theta)$};

\draw[arr] (jointlaw) -- (waiting);
\draw[arr] (jointlaw) -- (jump);


\node[mainbox,
      below=0.70cm of $(waiting)!0.5!(jump)$] (renewal)
{GENERAL RENEWAL STEP PROCESS
 $\xi[t]$};

\draw[arr] (waiting) -- (renewal);
\draw[arr] (jump) -- (renewal);

\node[mainbox, below=of renewal] (singleepoch)
{SINGLE-EPOCH STRUCTURE
$P_1(\Delta)$ and $p(\xi)\,\mathrm d\xi$\\
{\scriptsize correlated case:
$P_1(\Delta,\xi)=p_\Delta(\xi)\,P_1(\Delta)$}};

\draw[arr] (renewal) -- (singleepoch);

\draw[dashedarr]
(coupled.east) -- ++(0.3,0) |-
(singleepoch.east);


\node[routebox,
      below left=0.4cm and 0.35cm of singleepoch] (cumulant)
{CUMULANT ROUTE\\
multi-time statistics};

\node[routebox,
      below right=0.4cm and 0.35cm of singleepoch] (frozen)
{FROZEN-NOISE ROUTE\\
piecewise-deterministic flows};

\draw[arr] (singleepoch) -- (cumulant);
\draw[arr] (singleepoch) -- (frozen);


\node[routebox, line width=1.4pt, below=of cumulant] (decomposition)
{Exact $G$-cumulant decomposition\\
Boolean cumulants $b_n$\\
and renewal weights $D_\sigma$};

\node[routebox, below=of frozen] (flows)
{Frozen flows\\[1mm]
$\dot x=-C(x)+\epsilon\xi I(x)$};

\draw[arr] (cumulant) -- (decomposition);
\draw[arr] (frozen) -- (flows);


\node[resultbox, line width=1.4pt, below=of decomposition] (kernel)
{Reduced dynamics\\
memory kernel and relaxation};

\node[resultbox, below=of flows] (density)
{Stationary measures\\
support, singularities, tails;\\ strong events};

\draw[arr] (decomposition) -- (kernel);
\draw[arr] (flows) -- (density);


\node[resultbox, line width=1.4pt, below=of kernel] (moment)
{Long-lag moment structure\\[1mm]
$c_n\longrightarrow m_n$\\
{\scriptsize correlated case:
$m_n\longrightarrow m_n^{(\Delta)}$}};

\node[resultbox, below=of density] (fixedpoints)
{Long-epoch frozen structure\\[1mm]
fixed points and frozen rates\\
{\scriptsize weighted by $p_\Delta$ if correlated}};

\draw[arr] (kernel) -- (moment);
\draw[arr] (density) -- (fixedpoints);


\path let \p1=(moment.south), \p2=(fixedpoints.south), \p3=(singleepoch)
  in coordinate (mergetop) at (\x3,{min(\y1,\y2)-0.5cm});

\node[mergebox, anchor=north] (merge) at (mergetop)
{HEAVY-TAILED RENEWAL WITH FINITE MEAN:\\
THE TWO ROUTES MERGE\\[1mm]
$P_1$ controls cumulants, kernel, relaxation, singularities,
tails and strong events; amplitude--duration correlations are expected to
replace $p$ by the survival-selected law $p_\Delta$};

\draw[arr] (moment) |- (merge);
\draw[arr] (fixedpoints) |- (merge);


\node[comparisonbox, anchor=north east] (poisson)
at ($(merge.south)+(-0.25,-0.8)$)
{BOOLEAN CLASS\\[0.5mm]
Poissonian renewal, and exponential tails at long lags if
$\delta>\gamma_\psi$: kernel weighted by the Boolean L\'evy measure};

\node[comparisonbox, anchor=north west] (aging)
at ($(merge.south)+(0.25,-0.8)$)
{AGE-DEPENDENT CLASS\\[0.5mm]
infinite mean, $1<\mu<2$: coefficients depend on the age of the process};

\draw[thick, dashed] (merge.south) -- ++(0,-0.45) -| (poisson.north);
\draw[thick, dashed] (merge.south) -- ++(0,-0.45) -| (aging.north);
\node[font=\scriptsize, anchor=west, inner sep=1pt] at ($(merge.south)+(0.08,-0.2)$)
{contrasting universality classes};

\end{tikzpicture}%
}%
\caption{Program of the series of which this paper is the first. Under
amplitude--duration factorization, renewal step noise admits two
complementary descriptions: the cumulant route, which determines the memory
kernel and the relaxation, and the frozen-noise route, which determines the
stationary supports, singularities and tails and the statistics of strong
events. For heavy-tailed waiting times with finite mean, single-epoch
dominance makes the two descriptions merge asymptotically. Dashed boxes: the
other classes of waiting-time laws and the extension to correlated amplitudes
and durations. The boxes with thick frames are developed in the present
paper, for power-law waiting times; the relaxation is not treated here.}
\label{fig:conceptual_map}
\end{figure}
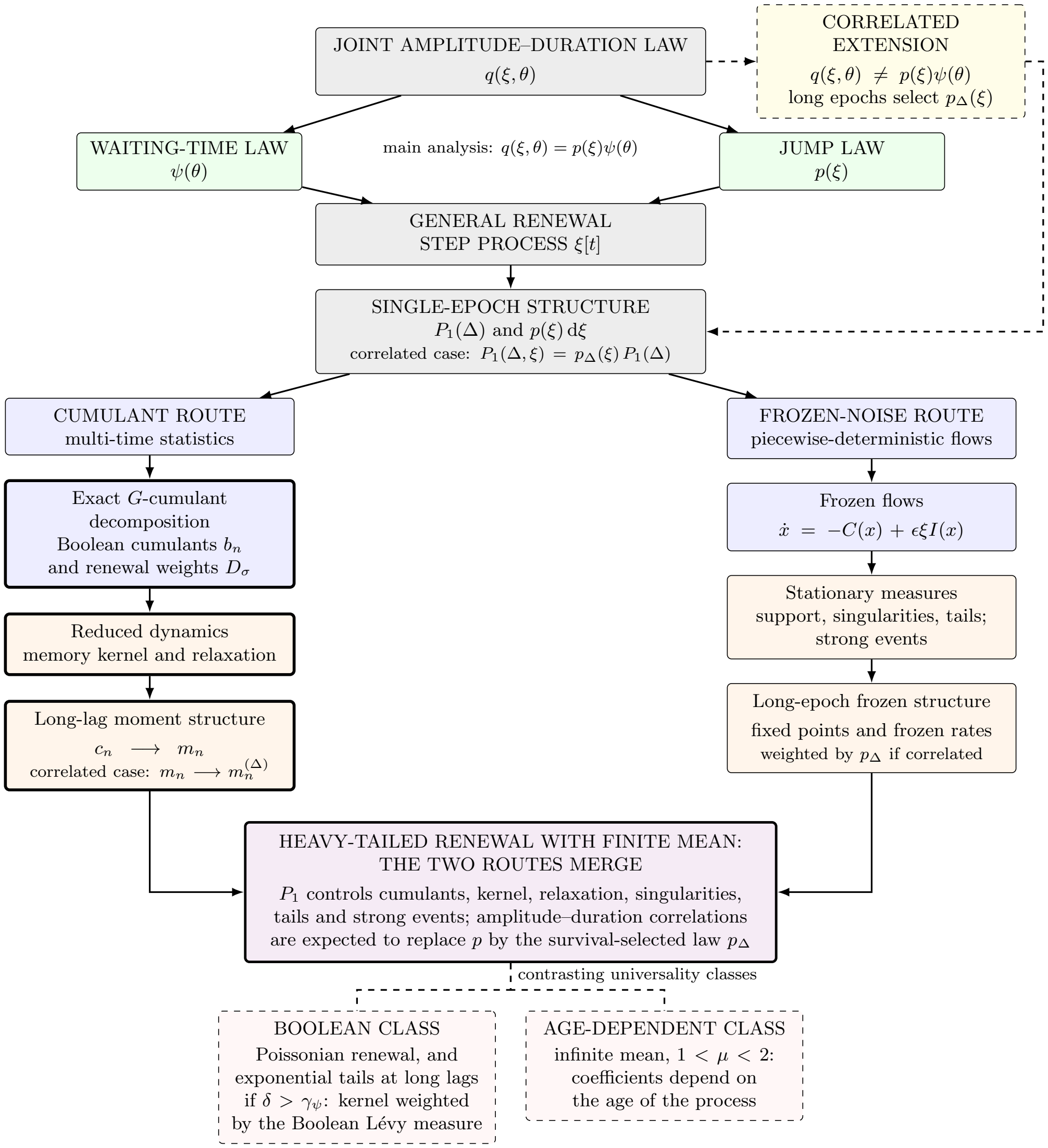

This paper is the first of a series devoted to systems driven by renewal
step noise with general waiting-time laws. Figure~\ref{fig:conceptual_map}
shows the program. The present paper develops the cumulant route for
power-law waiting times: the exact decomposition, the crossover to the
moments and the resulting long-lag kernel. Already here the two routes meet,
since the long-lag kernel is an average over the frozen flows weighted by the
jump law itself. The other boxes, which concern other classes of
waiting-time laws, the relaxation, the stationary densities and the
statistics of strong events, are the subject of separate papers; nothing in
the present one depends on them.

Statements labeled Theorem, Proposition, Lemma or Corollary are proved, in
the main text or in Appendix~\ref{app:proofs}. Result~\ref{prop:spectral},
which concerns the driven system, is a single-epoch prediction: it relies on
the dominance of long epochs at all the lags that contribute to the kernel,
and it is tested against exact calculations in Section~\ref{sec:limi}.

The paper is organized as follows. Section~\ref{sec:setting} introduces the
renewal step process, the $G$-cumulants and the memory kernel, the frozen
flows, the Poissonian kernel and the LIMI/CAM model.
Section~\ref{sec:decomposition} derives the exact decomposition and its
consequences for the low orders of the kernel. Section~\ref{sec:powerlaw}
proves the crossover to the moments for power-law waiting times,
Section~\ref{sec:ht_kernel} derives the long-lag kernel and the corresponding
master equation, and Section~\ref{sec:limi} tests them on the LIMI/CAM model.
Section~\ref{sec:discussion} discusses the results, and Section~\ref{sec:conclusions}
summarizes them. The proofs and the
numerical methods are collected in the appendices. Each section opens with a
short \emph{Summary}.

\section{Setting\label{sec:setting}}
\begin{summary}
The forcing is a sequence of epochs with random amplitudes and durations. The
key temporal quantity is the probability that one epoch covers a prescribed
observation span. This section fixes the notation, recalls the correlation and
$G$-cumulant results used below and the expansion of the memory kernel in
$G$-cumulants. The Poissonian case serves as the reference: there, each
$G$-cumulant is a Boolean cumulant of the jump law multiplied by a
single-epoch probability, and the memory kernel is an average of frozen-flow
kernels over the Boolean L\'evy measure of the jump law. Finally, the section
introduces the LIMI/CAM model, on which the results are tested.
\end{summary}

\subsection{Renewal step noise}
The process consists of successive renewal epochs indexed by $k$. Each epoch
carries an amplitude--duration pair $(\xi_k,\theta_k)$. Successive pairs are
independent and identically distributed according to a joint probability
density $q(\xi,\theta)$, normalized by
\begin{equation}
\int_{-\infty}^{\infty}\!\mathrm d\xi
\int_0^\infty\!\mathrm d\theta\,q(\xi,\theta)=1.
\label{eq:joint_law_normalization}
\end{equation}
Its marginals are
\begin{equation}
p(\xi)=\int_0^\infty q(\xi,\theta)\,\mathrm d\theta,
\qquad
\psi(\theta)=\int_{-\infty}^{\infty}q(\xi,\theta)\,\mathrm d\xi.
\label{eq:joint_law_marginals}
\end{equation}
Throughout the main analysis, amplitude and duration within the same epoch are
independent, so that
\begin{equation}
q(\xi,\theta)=p(\xi)\psi(\theta).
\label{eq:joint_law_factorized}
\end{equation}
For discrete or mixed jump laws, $p(\xi)$ and $q(\xi,\theta)$ are understood
as generalized densities containing Dirac delta functions. Probability
elements are therefore written as $p(\xi)\,\mathrm d\xi$ and
$q(\xi,\theta)\,\mathrm d\xi\,\mathrm d\theta$ throughout. This
factorization permits the exact separation between jump and renewal
statistics derived below.

We use the notation of Ref.~\cite{bblmCSF202}. The process starts with a
renewal at the initial time $t_0$, and the amplitude of the first epoch is
drawn from $p_0$; unless stated otherwise, $p_0=p$. The observation times are
ordered, $t_0\le t_1\le\dots\le t_n$; $u$ denotes renewal instants and
$\theta$ waiting times. The survival probability is
$\Psi(\theta)=\int_\theta^\infty\psi$, the renewal density (the rate function
of Ref.~\cite{bblmCSF202}) is $R(t)=\sum_{k\ge1}\psi_k(t)$, with $\psi_k$ the
$k$-fold convolution of $\psi$, and $\tilde R(t)=\delta(t)+R(t)$. Moments of
the jump law are denoted $m_k:=\overline{\xi^k}$, and throughout
$\overline{\xi}=0$. The mean waiting time
$\tau=\int_0^\infty\theta\psi(\theta)\,\mathrm d\theta$ is finite, for
regularly varying tails $\psi(\theta)\sim\theta^{-\mu}$, for $\mu>2$; for
$\mu=2$ it depends on the slowly varying factor. As a reference power-law density we use the Pareto
(also called Manneville~\cite{bblmCSF202}) form
\begin{equation}
\psi(\theta)=\frac{\mu-1}{T}\Big(\frac{T}{\theta+T}\Big)^{\mu},\qquad
\Psi(\theta)=\Big(\frac{T}{\theta+T}\Big)^{\mu-1},\qquad
\tau=\frac{T}{\mu-2}\quad(\mu>2).
\label{eq:manneville}
\end{equation}
The gaps between consecutive
observation times are $d_i:=t_{i+1}-t_i$, and $\Delta:=t_n-t_1$ is the total
span.

For $t_i\le t_j$ we denote by
\begin{equation}
P_1(t_i,t_j):=\Pr\big[\text{no renewal in }[t_i,t_j]\big]
\label{eq:P1}
\end{equation}
the probability that $t_i$ and $t_j$ lie in the same epoch. With $p_0=p$ it
coincides with the two-time normalized correlation function
$\langle t_i\,t_j\rangle$ of Ref.~\cite{bblmCSF202}, open on both sides: the
epoch containing $t_i$ starts at the last renewal before $t_i$, at any time
between $t_0$ and $t_i$, and ends at any time after $t_j$,
\[
P_1(t_i,t_j)=\langle t_i\,t_j\rangle
=\int_{t_0}^{t_i}\mathrm du\,\tilde R(u-t_0)\,\Psi(t_j-u).
\]
In the stationary regime ($t_1-t_0\to\infty$, $\tau<\infty$) it depends only
on the lag,
\begin{equation}
P_1(t_i,t_j)=P_1(t_j-t_i),\qquad
P_1(x)=\frac1\tau\int_x^\infty\Psi(y)\,\mathrm dy,
\label{eq:P1_stat}
\end{equation}
which for Eq.~\eqref{eq:manneville} gives exactly
$P_1(x)=[T/(x+T)]^{\mu-2}$. In this regime the
first renewal after any time has the forward-recurrence density
$f_e(x)=\Psi(x)/\tau$.
\subsection{Correlation functions and \texorpdfstring{$G$}{G}-cumulants}
Let $\mathcal I(n)$ be the set of interval partitions (compositions) of
$\{1,\dots,n\}$, and let $1_n$ be the single-block partition. For a given
$\pi\in\mathcal I(n)$, let $W_\pi(t_1,\dots,t_n)$ be the probability that the
grouping of the observation times into epochs is \emph{exactly} $\pi$. This is
precisely the concatenated correlation function introduced in
Ref.~\cite{bblmCSF202}: for $\pi=\{1,\dots,i_1\}\{i_1+1,\dots,i_2\}\cdots
\{i_{p-1}+1,\dots,n\}$,
\begin{equation}
\label{eq:W_conc}
W_\pi(t_1,\dots,t_n)=\langle t_1\,t_{i_1}\conc t_{i_1+1}\,t_{i_2}\conc\cdots
\conc t_{i_{p-1}+1}\,t_n\rangle ,
\end{equation}
and $W_{1_n}=\langle t_1\,t_n\rangle=P_1(t_1,t_n)$. The weights $W_\pi$ are
properties of the renewal process alone, and Ref.~\cite{bblmCSF202} gives them
as nested integrals of $\psi$, $\Psi$ and $R$. We do not need these lengthy
expressions in general: we use the weights through their combinations, with
explicit representations where needed [Eq.~\eqref{eq:A1234} at fourth
order]. The central result of
Ref.~\cite{bblmCSF202}, rewritten in the present notation, is
\begin{equation}
C_n(t_1,\dots,t_n):=\langle\xi(t_1)\cdots\xi(t_n)\rangle
=\sum_{\pi\in\mathcal I(n)}\Big[\prod_{B\in\pi}m_{|B|}\Big]\,W_\pi(t_1,\dots,t_n),
\label{eq:Cn_moments}
\end{equation}
valid for an arbitrary waiting-time density $\psi$, under the factorized
joint law~\eqref{eq:joint_law_factorized} and with $p_0=p$.

The $G$-cumulants were introduced in
Ref.~\cite{bbJSTAT4} by combining the standard moment--cumulant relation with
total time ordering. Exactly as for Boolean cumulants, in combinatorial terms, they are obtained from the
correlation functions by M\"obius inversion on $\mathcal I(n)$~\cite{bmBoolean},
\begin{equation}
\doubleangle{\xi(t_1)\cdots\xi(t_n)}^{(G)}
=\sum_{\zeta\in\mathcal I(n)}(-1)^{|\zeta|-1}\prod_{Z\in\zeta}C_{|Z|}(t_Z),
\label{eq:kappaG_def}
\end{equation}
where $t_Z$ denotes the observation times with indices in $Z$.
In the generalized-cumulant construction of Ref.~\cite{bbJSTAT4} the
cumulants are fixed by a cumulant map, which decides which partitions of the
times survive and with which weight. Different maps give the classical cumulants
and the free and monotone cumulants of non-commutative
probability~\cite{SpeicherWoroudi1997,HasebeSaigo2011,ahlvAM282,bMcum}. The
total time ordering requires the last time of each block to precede the first
time of the next. It therefore keeps exactly the interval partitions, each with
unit weight, and the $G$-cumulants are the Boolean cumulants of the
time-ordered family $\{\xi(t_1),\dots,\xi(t_n)\}$~\cite{bmBoolean,bMcum}: for
$t_1\le\dots\le t_n$, Eq.~\eqref{eq:kappaG_def} is the Boolean joint cumulant
of an ordered set of random variables,
\begin{equation}
b(X_1,\dots,X_n):=\sum_{\zeta\in\mathcal I(n)}(-1)^{|\zeta|-1}
\prod_{Z\in\zeta}\mathbb E\Big[\prod_{k\in Z}X_k\Big],
\label{eq:boolean_joint}
\end{equation}
evaluated on $\big(\xi(t_1),\dots,\xi(t_n)\big)$. Since the interval partitions depend on the order of the arguments, the time
ordering is part of this definition.  The Boolean cumulants of $p(\xi)$ are the special case of
equal arguments,
\begin{equation}
b_n=b(\xi,\dots,\xi)=\sum_{\zeta\in\mathcal I(n)}(-1)^{|\zeta|-1}\prod_{Z\in\zeta}m_{|Z|},
\qquad\text{equivalently}\qquad
m_n=\sum_{\sigma\in\mathcal I(n)}\prod_{B\in\sigma}b_{|B|}.
\label{eq:boolean_def}
\end{equation}

The $G$-cumulants are the coefficients of the memory kernel of the master
equation~\eqref{eq:ME}~\cite[Section~4.4.3]{bbJSTAT4},
\cite[Sections~2.1--2.2]{bmBoolean}:
\begin{align}
\mathcal G(t,u)
={}&
\sum_{n\geq2}\epsilon^n
\int_{u<u_2<\cdots<u_{n-1}<t}
\mathrm du_2\cdots\mathrm du_{n-1}\,
\doubleangle{
\xi(u)\xi(u_2)\cdots\xi(u_{n-1})\xi(t)
}^{(G)}
\nonumber\\
&\times
\mathcal L_I
e^{\mathcal L_a(t-u_{n-1})}
\mathcal L_I\cdots
e^{\mathcal L_a(u_2-u)}
\mathcal L_I .
\label{eq:kernel_G}
\end{align}
For $n=2$ there are no intermediate times, and the first-order term is absent
because $\overline\xi=0$. Each term is a time-ordered integral of the
operators $\mathcal L_I$, connected by the unperturbed evolution and weighted
by a multi-time $G$-cumulant of the forcing. For stationary forcing the kernel
depends only on the lag $\Delta=t-u$, and we write $\mathcal G(\Delta)$.

For exponential waiting times, $\psi(\theta)=e^{-\theta/\tau}/\tau$, the
$G$-cumulants are exactly~\cite{bmBoolean}
\begin{equation}
\doubleangle{\xi(t_1)\cdots\xi(t_n)}^{(G)}=b_n\,e^{-(t_n-t_1)/\tau}.
\label{eq:boolean_poisson}
\end{equation}
This is the non-trivial content of the Poissonian result: the multi-time
quantity~\eqref{eq:kappaG_def} reduces to the one-time Boolean cumulant $b_n$
of the jump law, multiplied by a survival factor. In what follows,
``Boolean structure'' always refers to this reduction, that is, to the
coefficients $b_n$ of $p(\xi)$.

\subsection{Frozen flows}
\label{sec:frozen_def}
For each value $\xi$ of the forcing, Eq.~\eqref{eq:SDE} defines the frozen
flow $\dot x=f_\xi(x):=-C(x)+\epsilon\xi I(x)$, with flow map $\phi^t_\xi$ and
generator $\mathcal L_\xi:=\mathcal L_a+\epsilon\xi\mathcal L_I$; during an
epoch with value $\xi$ the system follows $\phi^t_\xi$. No linearity of $C$ or
$I$ is assumed. We consider the \emph{stable regime}: for every $\xi$ in the
support of $p$, $f_\xi$ has a single fixed point $x^*(\xi)$, which attracts
every point of the region visited by the system and is hyperbolic, with rate
\begin{equation}
\lambda(\xi):=-f'_\xi\big(x^*(\xi)\big)>0 .
\label{eq:frozen_rate}
\end{equation}

\subsection{The Poissonian kernel\label{sec:poisson_kernel}}
For exponential waiting times the kernel~\eqref{eq:kernel_G}, with the
$G$-cumulants~\eqref{eq:boolean_poisson}, can be summed over
$n$~\cite{bmBoolean}. The result is best expressed through the Boolean L\'evy
measure of the jump law~\cite{bmBoolean}: for a law with finite variance there
is a unique positive generalized density $\varrho(x)$, of mass $b_2$, whose
moments, when finite, are the higher Boolean cumulants,
$b_{n+2}=\int_{-\infty}^{\infty}x^n\varrho(x)\,\mathrm dx$. With the frozen
generator $\mathcal L_x=\mathcal L_a+\epsilon x\mathcal L_I$ at the value $x$,
the kernel is, exactly and at all lags~\cite{bmBoolean},
\begin{equation}
\mathcal G(\Delta)=\epsilon^2\,e^{-\Delta/\tau}\,
\mathcal L_I\Big[\int_{-\infty}^{\infty}\varrho(x)e^{\mathcal L_x\Delta}\,\mathrm dx\Big]\mathcal L_I .
\label{eq:kernel_poisson_rho}
\end{equation}
Indeed, expanding the frozen propagator in a Dyson series in
$\epsilon x\mathcal L_I$,
\begin{equation}
e^{\mathcal L_x\Delta}=\sum_{k\ge0}(\epsilon x)^k\int_{\substack{d_i>0\\\sum_{i=0}^k d_i=\Delta}}
e^{\mathcal L_ad_0}\mathcal L_I e^{\mathcal L_ad_1}\cdots\mathcal L_I e^{\mathcal L_ad_k}
\prod_i\mathrm dd_i ,
\label{eq:dyson}
\end{equation}
sandwiching it between two factors $\epsilon\mathcal L_I$ and averaging $x^k$
over a measure $\omega$ gives a chain of the form~\eqref{eq:kernel_G}, with
$n=k+2$ couplings and coefficient $\int_{-\infty}^{\infty}x^k\omega(x)\,\mathrm dx$. For
$\omega=\varrho$ these coefficients are $b_{k+2}$, and with the factor
$e^{-\Delta/\tau}$ the chains are those of Eq.~\eqref{eq:kernel_G} with the
$G$-cumulants~\eqref{eq:boolean_poisson}, which gives
Eq.~\eqref{eq:kernel_poisson_rho}. The Poissonian kernel is thus an average of
frozen-flow kernels, but its measure is the Boolean L\'evy density, the
``once-stripped'' jump law~\cite{bmBoolean}, which is not the law of the
values held by the noise. For the symmetric Bernoulli law $\xi=\pm a$,
$\varrho=a^2\delta_0$: all the Boolean cumulants beyond the second vanish, and
Eq.~\eqref{eq:kernel_poisson_rho} is the exact second-order telegraph kernel,
with the free propagator $e^{\mathcal L_a\Delta}$. In Laplace space the
product with $e^{-\Delta/\tau}$ is a shift $s\to s+1/\tau$; this is what gives
the closed resummation of the Poissonian kernel through the Boolean generator
of $p(\xi)$~\cite{bmBoolean}.

\subsection{The LIMI/CAM model\label{sec:limi_model}}
The results are tested on the linear model with correlated additive and
multiplicative (CAM) forcing
\begin{equation}
\dot x=-\gamma x+\epsilon\,\xi[t]\,(1+\beta x),
\label{eq:limi}
\end{equation}
i.e.\ Eq.~\eqref{eq:SDE} with $C(x)=\gamma x$ and $I(x)=1+\beta x$, referred
to below as the LIMI/CAM model. Models of this form are widely used to
represent skewed and heavy-tailed atmospheric and oceanic
statistics~\cite{ssJC22,psC22,sAR101,spC25}, including ENSO models driven
by state-dependent stochastic
forcing~\cite{bGRL43,bcmmCHAOS28,bmCHAOS31,jltzGRL34,ljJAS67,bmCHAOS34}. If $\xi[t]$ represents episodic
forcing, such as westerly wind bursts, whose durations need not be
exponentially distributed, the model poses the closure problem in a concrete
form: does slowing down the resolved dynamics suppress the higher-order terms
of the kernel? Its frozen flows are linear,
$\phi^t_\xi(x)=x^*(\xi)+[x-x^*(\xi)]\,e^{-\lambda(\xi)t}$, with
\begin{equation}
x^*(\xi)=\frac{\epsilon\xi}{\lambda(\xi)},\qquad
\lambda(\xi)=\gamma-\epsilon\beta\xi .
\label{eq:limi_frozen}
\end{equation}
The system is in the stable regime if $\lambda(\xi)>0$ on the support of $p$,
i.e., for $\epsilon\beta>0$, if the support lies below
$\xi_{\rm c}=\gamma/(\epsilon\beta)$. Under amplitude--duration factorization
its stationary mean can be computed exactly for arbitrary jump and
waiting-time laws (Section~\ref{sec:limi}).

\section{Exact decomposition for arbitrary waiting times\label{sec:decomposition}}

\begin{summary}
For a general waiting-time law, a $G$-cumulant contains additional terms
beyond the Poissonian contribution. The exact decomposition derived here
writes it as a sum over interval partitions of the observation times, in
which the jump law enters only through products of its Boolean cumulants
$b_k$, and the waiting times only through renewal weights $D_\sigma$, the
Boolean joint cumulants of the no-renewal events of the blocks. The weights
identify precisely how non-Poissonian renewal couples otherwise separated
observation intervals; they vanish identically only for Poissonian renewal.
Inserted in the memory kernel, the decomposition shows that at second order
the kernel depends on the waiting-time law only through the single-epoch
probability, and that a second-order truncation is exact only for the
telegraph process.
\end{summary}

\begin{figure}[t]
\centering
\begin{tikzpicture}[x=1.15cm,y=0.85cm,
  blk/.style={fill=blue!12,draw=blue!55,rounded corners=2pt},
  ren/.style={red!75!black,line width=1.1pt},
  ax/.style={line width=0.6pt,-{Stealth[length=2mm]}}]
  \foreach \row/\lab/\res in {0/(a)/{$W_{\{12\}\{345\}\{678\}}$},
                              1/(b)/{$W_{\{12345\}\{678\}}$},
                              2/(c)/{$W_{1_8}=P_1(t_1,t_8)$},
                              3/(d)/{excluded}}{
    \pgfmathsetmacro{\y}{-1.25*\row}
    \fill[blk] (1.2-0.12,\y-0.22) rectangle (2.0+0.12,\y+0.22);
    \fill[blk] (3.3-0.12,\y-0.22) rectangle (4.5+0.12,\y+0.22);
    \fill[blk] (5.9-0.12,\y-0.22) rectangle (7.1+0.12,\y+0.22);
    \draw[ax] (-0.3,\y) -- (8.1,\y);
    \foreach \x/\n in {0.0/0,1.2/1,2.0/2,3.3/3,3.9/4,4.5/5,5.9/6,6.5/7,7.1/8}{
      \draw (\x,\y-0.1) -- (\x,\y+0.1);
      \node[below=3pt,font=\scriptsize] at (\x,\y-0.08) {$t_{\n}$};}
    \node[left,font=\small] at (-0.35,\y) {\lab};
    \node[right,font=\scriptsize,align=left] at (8.2,\y) {\res};
  }
  \foreach \x in {0.6,2.5,2.9,5.2,7.7} \draw[ren] (\x,-0.32) -- (\x,0.32);
  \foreach \x in {0.4,5.4,7.6} \draw[ren] (\x,-1.25-0.32) -- (\x,-1.25+0.32);
  \foreach \x in {0.8,7.8} \draw[ren] (\x,-2.5-0.32) -- (\x,-2.5+0.32);
  \foreach \x in {0.5,2.6,3.6,5.3} \draw[ren] (\x,-3.75-0.32) -- (\x,-3.75+0.32);
  \node[red!75!black,font=\bfseries] at (3.6,-3.75+0.52) {$\times$};
  \foreach \a/\b/\l in {1.2/2.0/{B_1},3.3/4.5/{B_2},5.9/7.1/{B_3}}
    \node[font=\scriptsize,blue!55!black] at ({(\a+\b)/2},0.5) {$\l$};
\end{tikzpicture}
\caption{Realizations of the renewal process and the probability $A_\sigma$
of Eq.~\eqref{eq:A_sigma}, for $n=8$ and $\sigma=\{12\}\{345\}\{678\}$.
Shaded boxes: the blocks $B_1,B_2,B_3$ of $\sigma$; vertical bars: renewal
instants. A realization contributes to $A_\sigma$ if no renewal falls inside a
block, whatever happens before $t_1$, after $t_8$ and in the gaps between
blocks (rows a--c). The interior times $t_4$ and $t_7$ add no constraint: a
block lies in one epoch if and only if its first and last times do. The
resulting grouping $\pi$ of the times into epochs is coarser than or
equal to $\sigma$; on the right, the probability $W_\pi$ to which each
realization contributes. Summing over these groupings gives
$A_\sigma=\sum_{\pi\ge\sigma}W_\pi$. Row (c), a single epoch covering
$[t_1,t_8]$, contributes $P_1(t_1,t_8)$, so that $A_\sigma\ge P_1(t_1,t_8)$.
Row (d): a renewal between $t_3$ and $t_4$, inside $B_2$, excludes the
realization.}
\label{fig:Asigma}
\end{figure}
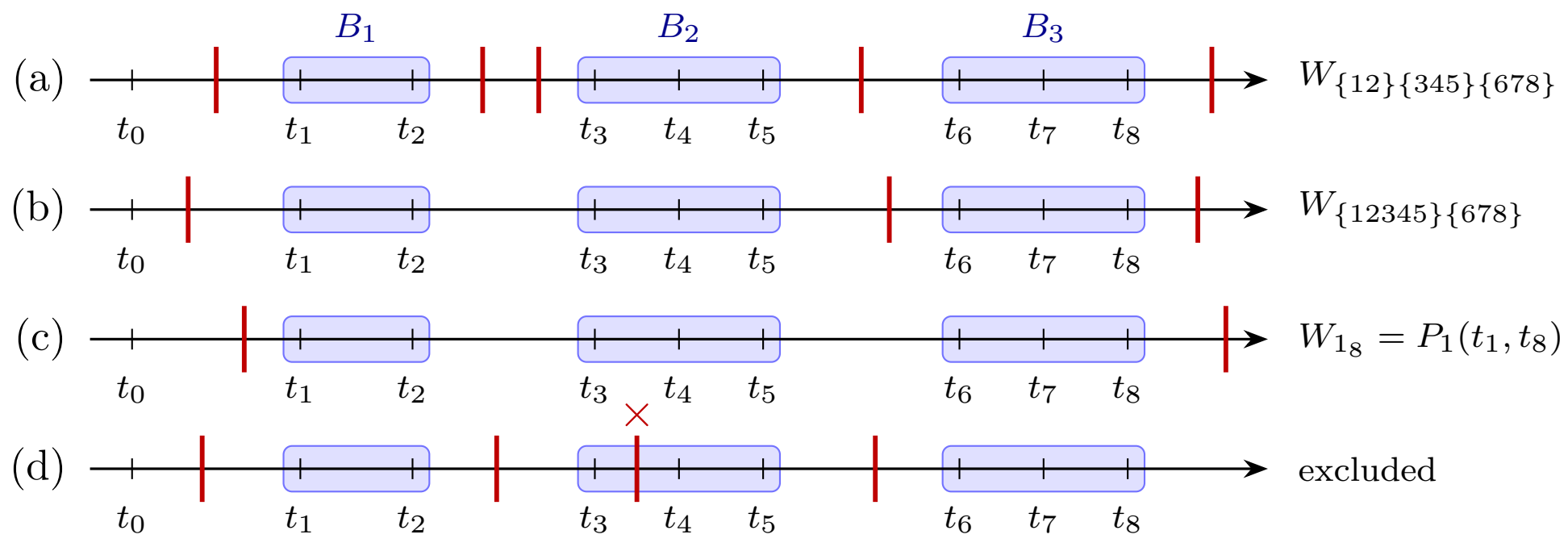

For an interval partition $\sigma\in\mathcal I(n)$ and each block $B$ of
$\sigma$, let $E_B$ denote the event that no renewal occurs in
$[t_{\min B},t_{\max B}]$, and let
\begin{equation}
A_\sigma(t_1,\dots,t_n):=\Pr\Big[\bigcap_{B\in\sigma}E_B\Big]
\label{eq:A_sigma}
\end{equation}
be the probability that no renewal occurs inside any block of $\sigma$,
whatever happens in the gaps between blocks (Fig.~\ref{fig:Asigma}). Singleton
blocks impose no constraint, and $A_{1_n}=P_1(t_1,t_n)$. In the bracket
notation of Ref.~\cite{bblmCSF202}, $A_\sigma$ is the sum of the
concatenations in which each junction between consecutive blocks of $\sigma$
is either kept or removed, i.e.\ $A_\sigma=\sum_{\pi\ge\sigma}W_\pi$, where
$\pi\ge\sigma$ means that $\pi$ is obtained from $\sigma$ by merging
consecutive blocks; for instance
$A_{\{12\}\{34\}}=\langle t_1\,t_4\rangle+\langle t_1\,t_2\conc t_3\,t_4\rangle$.

\paragraph{Correlation functions in Boolean form.} Re-expanding the jump
moments in Boolean cumulants rewrites the correlation
function~\eqref{eq:Cn_moments}, for an arbitrary waiting-time density and
under the factorized joint law $q(\xi,\theta)=p(\xi)\psi(\theta)$, as
\begin{equation}
C_n(t_1,\dots,t_n)=\sum_{\sigma\in\mathcal I(n)}
\Big[\prod_{B\in\sigma}b_{|B|}\Big]\,A_\sigma(t_1,\dots,t_n).
\label{eq:Cn_general}
\end{equation}
Indeed, the indices of each block $B$ of $\pi$ in Eq.~\eqref{eq:Cn_moments}
are consecutive, so that the moment--Boolean relation~\eqref{eq:boolean_def}
can be applied to $B$ alone,
$m_{|B|}=\sum_{\sigma_B\in\mathcal I(B)}\prod_{B'\in\sigma_B}b_{|B'|}$.
Expanding the product over the blocks of $\pi$, the blocks of all the
$\sigma_B$ form an interval partition $\sigma$ that refines $\pi$
($\sigma\le\pi$), and every $\sigma\le\pi$ arises exactly once; for
$\pi=\{12\}\{34\}$, for instance, $m_2m_2=(b_2+b_1^2)^2$ produces the four
refinements $\{12\}\{34\}$, $\{1\}\{2\}\{34\}$, $\{12\}\{3\}\{4\}$ and
$\{1\}\{2\}\{3\}\{4\}$. Exchanging the order of summation and using
$\sum_{\pi\ge\sigma}W_\pi=A_\sigma$ gives Eq.~\eqref{eq:Cn_general}.

Each grouping of the observation times into blocks of consecutive times thus
contributes two factors: a product of Boolean cumulants of the jump law, one
for each block, and the probability $A_\sigma$ that every block lies within a
single epoch. The jump law enters only through the $b_k$, the renewal process
only through the $A_\sigma$. Compared with Eq.~\eqref{eq:Cn_moments}, the
moments $m_{|B|}$ are replaced by Boolean cumulants, and the concatenation
$W_\pi$ by the sum $A_\sigma$ of all the concatenations in which no block of
$\sigma$ is split. Applying the Boolean inversion to
Eq.~\eqref{eq:Cn_general} gives the central result of this section.
\begin{proposition}[Exact jump--renewal separation]
\label{prop:kappaG}
For an arbitrary waiting-time density, under the factorized joint law
$q(\xi,\theta)=p(\xi)\psi(\theta)$ and with $p_0=p$,
\begin{equation}
\doubleangle{\xi(t_1)\cdots\xi(t_n)}^{(G)}
=\sum_{\sigma\in\mathcal I(n)}\Big[\prod_{B\in\sigma}b_{|B|}\Big]\,
D_\sigma(t_1,\dots,t_n),
\qquad
D_\sigma:=\sum_{\substack{\zeta\in\mathcal I(n)\\ \zeta\ge\sigma}}
(-1)^{|\zeta|-1}\prod_{Z\in\zeta}A_{\sigma|_Z},
\label{eq:kappaG_general}
\end{equation}
where $\sigma|_Z$ is the restriction of $\sigma$ to the block $Z$ of $\zeta$,
and $A_{\sigma|_Z}$ is evaluated at the times $t_Z$. Equivalently, if
$\sigma=\{B_1,\dots,B_p\}$ with blocks in increasing order and
$\chi_k:=\mathbf 1_{E_{B_k}}$,
\begin{equation}
D_\sigma=b(\chi_1,\dots,\chi_p),
\label{eq:D_boolean}
\end{equation}
the Boolean joint cumulant~\eqref{eq:boolean_joint} of the ordered
no-renewal indicators.
\end{proposition}

\begin{inwords}
The factorization of amplitude and duration induces an exact separation
between the jump statistics and the renewal statistics. The coefficients
$b_{|B|}$ depend only on $p(\xi)$, whereas the weights $D_\sigma$ depend only on
the renewal process, on the observation times and, before stationarity is
reached, on $t_0$. The weights $D_\sigma$ play for the $G$-cumulants, which are
the coefficients of the memory kernel, the role that the probabilities
$A_\sigma$ play for the correlation functions.
The weight $D_\sigma$ is the connected part of $A_\sigma$, in the Boolean sense: what remains after removing the products of probabilities referring to separate groups of consecutive blocks. A non-zero $D_\sigma$ with
$\sigma\ne1_n$ therefore measures how strongly the waiting-time law couples
observation intervals separated by gaps. When the no-renewal events of
different blocks are independent, as for Poissonian renewal, only the
single-block term $b_n\,P_1(t_1,t_n)$ survives.
\end{inwords}
\begin{proof}
Insert Eq.~\eqref{eq:Cn_general}, applied to each block $Z$ of $\zeta$, into
Eq.~\eqref{eq:kappaG_def}. A choice of $\sigma_Z\in\mathcal I(Z)$ for every
$Z\in\zeta$ is the same as an interval partition $\sigma\le\zeta$, and
exchanging the sums gives Eq.~\eqref{eq:kappaG_general}. The coarsenings
$\zeta\ge\sigma$ are in bijection with the interval partitions of the $p$
blocks of $\sigma$, and
$A_{\sigma|_Z}=\Pr\big[\bigcap_{B_k\subseteq Z}E_{B_k}\big]
=\mathbb E\big[\prod_{k:\,B_k\subseteq Z}\chi_k\big]$, which gives
Eq.~\eqref{eq:D_boolean}.
\end{proof}

Three simple properties of the weights are used below. First,
$D_{1_n}=P_1(t_1,t_n)$ is the only weight that multiplies $b_n$. Second, since
$b_1=\overline\xi=0$, partitions with singleton blocks do not contribute, and
only $\sigma$ with all blocks of size $\ge2$
matter.\footnote{For $\overline\xi\ne0$ one may check that $D_\sigma=0$
whenever the first or the last block of $\sigma$ is a singleton, because a
Boolean cumulant with a constant first or last argument vanishes. Interior
singleton blocks, however, give non-zero weights in general.} Third, when all
the temporal gaps vanish, $d_i\to0$, every $A_\sigma$ tends to one, and for
$\sigma\ne1_n$ we have $D_\sigma\to\sum_{\zeta\ge\sigma}(-1)^{|\zeta|-1}=0$,
because the coarsenings of $\sigma$ correspond to the subsets of its cuts.
Hence
\begin{equation}
\doubleangle{\xi(t_1)\cdots\xi(t_n)}^{(G)}\to b_n\qquad(d_i\to0):
\label{eq:short_lag}
\end{equation}
when all the observation times coincide, the $G$-cumulant reduces to the
Boolean cumulant of a single value, for every waiting-time law.

If, finally, the no-renewal events on the blocks are independent,
$A_{\sigma|_Z}=\prod_{B\subseteq Z}\Pr[E_B]$, the product
$\prod_{Z\in\zeta}A_{\sigma|_Z}$ does not depend on $\zeta$, and $D_\sigma=0$
for $\sigma\ne1_n$. This is the case for Poissonian renewal, for which
Eq.~\eqref{eq:kappaG_general} reduces to Eq.~\eqref{eq:boolean_poisson}. The
converse also holds.

\begin{corollary}[Uniqueness of the exact Boolean class]
\label{cor:poisson}
Let the renewal step process be stationary, with statistically independent
amplitudes and durations. Then
$\doubleangle{\xi(t_1)\cdots\xi(t_n)}^{(G)}=b_n\,P_1(t_n-t_1)$ for every
symmetric jump law, every $n$ and all times if and only if the waiting times
are exponential.
\end{corollary}

\begin{proof}
Sufficiency is Eq.~\eqref{eq:boolean_poisson}. Conversely, for $n=4$ and a
non-degenerate symmetric jump law, Eq.~\eqref{eq:kappaG_general} gives
$b_2^2\,D_{\{12\}\{34\}}$, with $b_2>0$, as the only additional term, so
$D_{\{12\}\{34\}}=A_{\{12\}\{34\}}-P_1(d_1)P_1(d_3)$ must vanish for all
$d_1,d_2,d_3$. As $d_2\to0$ the two blocks join into $[t_1,t_4]$, so that
$A_{\{12\}\{34\}}\to P_1(d_1+d_3)$ and $P_1(d_1+d_3)=P_1(d_1)P_1(d_3)$. Hence
$P_1$ is exponential, and so is $\psi$, since $\psi=\tau\,P_1''$ by
Eq.~\eqref{eq:P1_stat}.
\end{proof}

Exponential waiting times are thus the only ones for which the $G$-cumulants
have the Boolean form at all orders and all times. Any other waiting-time law
already leaves a trace at fourth order, through the weight that couples the
first and the last pair of observation times.

\begin{remark}[Atom surrogates]\label{rem:atoms}
Since the weights $D_\sigma$ do not depend on $p(\xi)$, replacing $p$ with its
$N$-point Gauss quadrature, which exists if $p$ has finite moments up to order
$2N-1$ and at least $N$ support points, and leaves $b_k$ unchanged for
$k\le2N-1$, preserves all the correlation functions and $G$-cumulants through
order $2N-1$, for any waiting-time law independent of the jump amplitudes.
This is an immediate proof of the correlation-preservation property of
Ref.~\cite{bmBoolean}.
\end{remark}

\subsection{Effective coefficients and the fourth-order index}

\begin{summary}
To compare different renewal laws, we normalize each $G$-cumulant by the
probability that one epoch covers the full observation span. The resulting
effective coefficients equal the Boolean cumulants in the Poissonian case and
can approach the ordinary moments at long lags for power-law renewal. A
fourth-order index, independent of the jump law, provides the lowest-order
diagnostic of this crossover.
\end{summary}

Normalizing by the single-epoch probability defines effective per-block
coefficients,
\begin{equation}
c_n(t_1,\dots,t_n):=
\frac{\doubleangle{\xi(t_1)\cdots\xi(t_n)}^{(G)}}{P_1(t_1,t_n)}
=b_n+\sum_{\substack{\sigma\in\mathcal I(n)\\ \sigma\ne1_n}}
\Big[\prod_{B\in\sigma}b_{|B|}\Big]\,r_\sigma,
\qquad r_\sigma:=\frac{D_\sigma}{P_1(t_1,t_n)} .
\label{eq:c_n}
\end{equation}
For Poissonian renewal $c_n\equiv b_n$. If every relevant $r_\sigma$ equals one,
then $c_n=\sum_\sigma\prod_Bb_{|B|}=m_n$. Where a given renewal process sits
between these two structures is measured by the normalized deviation
\begin{equation}
\rho_n:=\frac{c_n-b_n}{m_n-b_n}
=\frac{\sum_{\sigma\ne1_n}\big[\prod_{B\in\sigma}b_{|B|}\big]\,r_\sigma}
{\sum_{\sigma\ne1_n}\prod_{B\in\sigma}b_{|B|}},
\label{eq:rho_n}
\end{equation}
which is $0$ in the Boolean case and $1$ in the moment case, for every even
$n$, for which $m_n\ne b_n$. For a symmetric
jump law the odd Boolean cumulants vanish and the even ones are non-negative,
$b_{2k}=\int_{-\infty}^{\infty}x^{2k-2}\varrho(x)\,\mathrm dx$ with $\varrho$ the Boolean L\'evy measure
of Ref.~\cite{bmBoolean} (Section~\ref{sec:poisson_kernel}). Hence $\rho_n$ is an average of the weights
$r_\sigma$ with non-negative coefficients.
In general it depends on $p(\xi)$ through these coefficients: at $n=6$, for
instance, three weights contribute for a symmetric law, with coefficients
$b_2b_4$, $b_2b_4$ and $b_2^3$. Fourth order is special.
The only singleton-free partition other than $1_4$ is $\{12\}\{34\}$, so
$m_4-b_4=b_2^2$ and
\begin{equation}
\rho_4=r_{\{12\}\{34\}}
=\frac{A_{\{12\}\{34\}}-P_1(t_1,t_2)\,P_1(t_3,t_4)}{P_1(t_1,t_4)}
\label{eq:rho4}
\end{equation}
does not depend on $p(\xi)$. It is the lowest-order measure of the departure
from the Boolean structure, it is a property of the renewal process alone,
and, by Eq.~\eqref{eq:kappa4} below, it is the waiting-time part of the
fourth-order coefficient of the memory kernel.

 In the stationary regime,
\begin{equation}
A_{\{12\}\{34\}}=P_1(\Delta)
+\int_{t_2}^{t_3}\!\mathrm du\,f_e(u-t_1)
\int_u^{t_3}\!\mathrm du'\,\tilde R(u'-u)\,\Psi(t_4-u').
\label{eq:A1234}
\end{equation}
The first term is the probability of a single epoch over $[t_1,t_4]$. In the
second, the first renewal after $t_1$ occurs at $u\in(t_2,t_3)$, the last
renewal before $t_3$ at $u'\ge u$ (with $u'=u$ accounted for by the $\delta$
in $\tilde R$), and the following waiting time exceeds $t_4-u'$.

\subsection{Consequences for the memory kernel\label{sec:kernel_low}}

The $n$-th term of the memory kernel~\eqref{eq:kernel_G} of the master
equation~\eqref{eq:ME} contains the $n$-th $G$-cumulant of the noise, and
Proposition~\ref{prop:kappaG} gives these coefficients exactly for any
$\psi$. We write the resulting kernel and master equation for
Eq.~\eqref{eq:SDE} explicitly up to fourth order in $\epsilon$; the long-lag
kernel for power-law waiting times is derived at all orders in
Section~\ref{sec:ht_kernel}.

\paragraph{Second order.} For $n=2$ the only interval partition other than
$1_2$ is $\{1\}\{2\}$, whose weight vanishes, since singleton blocks impose
no constraint and $A_{\{1\}\{2\}}=1$. Hence, for every $\psi$, under the
factorized joint law, and in fact for any mean of $\xi$,
\begin{equation}
\doubleangle{\xi(t_1)\xi(t_2)}^{(G)}=b_2\,P_1(t_1,t_2).
\label{eq:kappa2}
\end{equation}
Inserted in Eq.~\eqref{eq:kernel_G}, with $\mathcal L_a\rho=\partial_x(C\rho)$
and $\mathcal L_I\rho=-\partial_x(I\rho)$, it gives the master equation for
Eq.~\eqref{eq:SDE} at second order,
\begin{equation}
\partial_tP(x\pv t)=\partial_x\big[C(x)P(x\pv t)\big]
+\epsilon^2b_2\int_{t_0}^t\mathrm du\,P_1(u,t)\,
\partial_x\Big\{I(x)\,e^{\mathcal L_a(t-u)}\,\partial_x\big[I(x)P(x\pv u)\big]\Big\}
+O(\epsilon^3).
\label{eq:ME2}
\end{equation}
The second-order kernel depends on the waiting-time law only through the
single-epoch probability, in agreement with the universality of the two-time
correlation function~\cite{bblmCSF196}.

\paragraph{Third and fourth order.} The only singleton-free interval
partition of three times is $1_3$, so that
$\doubleangle{\xi(t_1)\xi(t_2)\xi(t_3)}^{(G)}=b_3\,P_1(t_1,t_3)$, which
vanishes for a symmetric jump law. Since $\overline\xi=0$, the only weights
that contribute at fourth order are $D_{1_4}$ and $D_{\{12\}\{34\}}$, and
\begin{equation}
\doubleangle{\xi(t_1)\cdots\xi(t_4)}^{(G)}
=P_1(t_1,t_4)\big[b_4+b_2^2\,\rho_4\big].
\label{eq:kappa4}
\end{equation}
Collecting the terms, the kernel of the master equation for
Eq.~\eqref{eq:SDE} is
\begin{align}
&\mathcal G(t,u)=P_1(u,t)\Big\{\epsilon^2b_2\,\mathcal L_Ie^{\mathcal L_a(t-u)}\mathcal L_I
+\epsilon^3b_3\int_u^t\mathrm du_2\,
\mathcal L_Ie^{\mathcal L_a(t-u_2)}\mathcal L_Ie^{\mathcal L_a(u_2-u)}\mathcal L_I
\nonumber\\
&+\epsilon^4\int_{u<u_2<u_3<t}\mathrm du_2\,\mathrm du_3\,
\big[b_4+b_2^2\,\rho_4(u,u_2,u_3,t)\big]\,
\mathcal L_Ie^{\mathcal L_a(t-u_3)}\mathcal L_Ie^{\mathcal L_a(u_3-u_2)}
\mathcal L_Ie^{\mathcal L_a(u_2-u)}\mathcal L_I\Big\}\nonumber \\
&+O(\epsilon^5).
\label{eq:kernel_low}
\end{align}
Up to third order the waiting-time law enters only through $P_1$; at fourth
order it enters also through the index $\rho_4$. In the Poissonian case the
ratio between the fourth- and second-order coefficients is the purely
statistical factor $b_4/b_2$ that controls the error of a second-order
truncation~\cite{bmBoolean}. Here it becomes $(b_4+b_2^2\rho_4)/b_2$. For
power-law waiting times it grows from $b_4/b_2$ at short lags
[Eq.~\eqref{eq:short_lag}] to $m_4/b_2$ at long lags
(Theorem~\ref{thm:crossover}).

\begin{corollary}[Unique exact second-order closure]
\label{cor:closure}
For stationary step-renewal noise with statistically independent amplitudes
and durations, the fourth $G$-cumulant vanishes identically if and only if
the jump law is symmetric Bernoulli \emph{and} the waiting times are
exponential. For this process, the telegraph process, all the $G$-cumulants
beyond the second vanish, and the memory kernel closes exactly at second
order. In particular, dichotomous noise with non-exponential waiting times
does not admit an exact second-order closure.
\end{corollary}

For the telegraph process, i.e.\ symmetric dichotomous Markov noise
$\xi=\pm a$ with correlation time $\tau$, $P_1(u,t)=e^{-(t-u)/\tau}$ at all
times, and Eq.~\eqref{eq:ME2} holds without remainder:
\begin{equation}
\partial_tP(x\pv t)=\partial_x\big[C(x)P(x\pv t)\big]
+\epsilon^2a^2\int_{t_0}^t\mathrm du\,e^{-(t-u)/\tau}\,
\partial_x\Big\{I(x)\,e^{\mathcal L_a(t-u)}\,\partial_x\big[I(x)P(x\pv u)\big]\Big\}.
\label{eq:ME_telegraph}
\end{equation}
This is, in memory-kernel form, the known exact equation for systems driven
by dichotomous Markov noise. It follows by eliminating $\langle\xi\rho\rangle$ from the closed pair
of equations for $\langle\rho\rangle$ and $\langle\xi\rho\rangle$ given by the
differentiation formula of Shapiro and Loginov~\cite{slPhysicaA91} (see also
Refs.~\cite{Horsthemke1984,bIJMPB20}). In the language of generalized
cumulants it expresses the $G$-Gaussian character of the telegraph
process~\cite{bCSF159}, and it is the case $\varrho=a^2\delta_0$ of the
Boolean resummation~\eqref{eq:kernel_poisson_rho}~\cite{bmBoolean}. The
sufficiency part of the corollary is therefore known. Its new content is the
converse: among renewal step processes with independent amplitudes and
durations, no other combination of jump and waiting-time laws closes the
kernel exactly at second order.

\begin{proof}
By Eq.~\eqref{eq:kappa4}, the fourth $G$-cumulant vanishes identically if and
only if $b_4+b_2^2\rho_4\equiv0$. At short lags $\rho_4\to0$
[Eq.~\eqref{eq:short_lag}], so $b_4=0$. Since $\overline\xi=0$,
$b_4=m_4-m_2^2$ is the variance of $\xi^2$, which vanishes only if $|\xi|$ is
constant, i.e.\ for the symmetric Bernoulli law. Then $b_2^2\rho_4\equiv0$,
i.e.\ $D_{\{12\}\{34\}}\equiv0$, and the proof of
Corollary~\ref{cor:poisson} gives exponential waiting times. Conversely, for
the symmetric Bernoulli law $\xi=\pm a$ all the Boolean cumulants beyond the
second vanish, since $b_{n+2}=\int x^n\varrho(x)\,\mathrm dx$ with
$\varrho=a^2\delta_0$ (Section~\ref{sec:poisson_kernel}), and for exponential waiting times
Eq.~\eqref{eq:boolean_poisson} gives $G$-cumulants proportional to $b_n$: all
of them vanish beyond the second.
\end{proof}

With any other waiting-time law, dichotomous noise has a fourth $G$-cumulant
that does not vanish identically, and for power-law waiting times the
violation is maximal at long lags: $b_4=0$, but $c_4\to b_2^2=m_4$, the moment
value.
\section{Power-law waiting times: the crossover to moments\label{sec:powerlaw}}

\begin{summary}
For stationary renewal with regularly varying tails and finite mean, a long
observation span is most likely covered by one exceptionally long epoch. This
single-epoch dominance makes the weights of all the partitions without
singleton blocks approach the single-epoch probability, and reconstructs the
ordinary moments of the jump density, at all orders. Its physical
implication, derived in Section~\ref{sec:ht_kernel}, is that the long-lag
memory kernel becomes an average of frozen-flow kernels, weighted by
$\xi^2p(\xi)$ instead of the Boolean L\'evy measure that weights the
Poissonian kernel.
\end{summary}

Let $\sigma=\{B_1,\dots,B_p\}\in\mathcal I(n)$, with blocks in increasing
order, and let $\ell_k:=t_{\max B_k}-t_{\min B_k}$ be the span of the block
$B_k$ and $\ell_*:=\min_k\ell_k$.

\paragraph{Two-epoch bounds.} A single epoch that covers the whole span
$[t_1,t_n]$ satisfies all the block constraints at once, so that $A_\sigma$ is
never smaller than $P_1(\Delta)$. The excess of $A_\sigma$ over $P_1(\Delta)$
comes only from histories in which at least two different epochs each cover a
whole block. $D_\sigma$ differs from $A_\sigma$ by products of at least two
factors $A_{\sigma|_Z}$, Eq.~\eqref{eq:kappaG_general}. For a stationary
renewal process with $\tau<\infty$ and bounded renewal density,
$\|R\|_\infty:=\sup_{t>0}R(t)<\infty$, every $\sigma\in\mathcal I(n)$ with
$p\ge2$ blocks satisfies, as shown in Appendix~\ref{app:two_plateau},
\begin{align}
0\le A_\sigma-P_1(\Delta)&\le P_1(\ell_1)\big[\Psi(\ell_p)
+\tau\|R\|_\infty\,P_1(\ell_p)\big],
\label{eq:lemma_A}\\
\big|D_\sigma-P_1(\Delta)\big|&\le P_1(\ell_1)\big[\Psi(\ell_p)
+\tau\|R\|_\infty\,P_1(\ell_p)\big]+\big(2^{p-1}-1\big)\,P_1(\ell_*)^2 .
\label{eq:lemma_D}
\end{align}
Each term on the right-hand sides is a product of two factors, each the
probability that an epoch covers a block. This is the precise sense in which
the corrections to the single-epoch term are two-epoch effects. The bounds
assume nothing on the tail of the waiting times. They only require a bounded
renewal density, which holds whenever $\psi$ is directly Riemann integrable,
in particular when $\psi$ is bounded, piecewise continuous and eventually
non-increasing. The bounds are useful only when all the blocks are long.
Singleton blocks ($\ell_k=0$) are therefore excluded, at no cost, since they
do not contribute for $\overline\xi=0$.

\begin{theorem}[All-order crossover from Boolean cumulants to moments]
\label{thm:crossover}
Let $\psi$ be directly Riemann integrable, with a regularly varying tail, i.e.,
\begin{equation}
\Psi(\theta)=\theta^{-(\mu-1)}L(\theta),\qquad L\ \text{slowly varying},
\label{eq:Psi_rv}
\end{equation}
and either $\mu>2$, or $\mu=2$ with $\tau<\infty$. In the stationary regime,
let the span $\Delta=t_n-t_1$ grow with fixed gap fractions
$a_i:=d_i/\Delta>0$. Then, for every $\sigma\in\mathcal I(n)$ whose blocks all
have size $\ge2$,
\begin{equation}
D_\sigma=P_1(\Delta)\Big[1+O\big(P_1(\Delta)\big)\Big],
\qquad\Delta\to\infty .
\label{eq:crossover}
\end{equation}
Consequently $c_n=m_n+O\big(P_1(\Delta)\big)$ and
\begin{equation}
\doubleangle{\xi(t_1)\cdots\xi(t_n)}^{(G)}
=m_n\,P_1(\Delta)+O\big(P_1(\Delta)^2\big),\qquad
C_n(t_1,\dots,t_n)=m_n\,P_1(\Delta)+O\big(P_1(\Delta)^2\big),
\label{eq:universal}
\end{equation}
with $P_1(\Delta)\simeq L(\Delta)\,\Delta^{-(\mu-2)}/[(\mu-2)\tau]$ for
$\mu>2$.
\end{theorem}

\noindent Here $O(P_1)$ means that $|D_\sigma/P_1(\Delta)-1|\le
K_\sigma P_1(\Delta)$ for $\Delta$ large enough, with $K_\sigma$ depending on
$\sigma$ and on the gap fractions but not on $\Delta$. The $O(P_1^2)$ terms
are meant in the same way. Since $P_1(\Delta)\to0$, the relative correction
vanishes at least as fast as the single-epoch probability itself. The
asymptotic form of $P_1$ follows from Karamata's theorem. For the
Pareto/Manneville density~\eqref{eq:manneville},
$P_1(\Delta)=[T/(\Delta+T)]^{\mu-2}$ exactly, and the relative corrections in
Eq.~\eqref{eq:crossover} are $O\big((T/\Delta)^{\mu-2}\big)$.

\begin{inwords}
For power-law tails with a finite mean, a long span is much more likely to be
covered by one epoch than by two epochs, each covering a whole block. All the
weights $D_\sigma$ of interval partitions without singleton blocks then
approach the same value, the single-epoch probability $P_1(\Delta)$, with a
relative error of order $P_1(\Delta)$ itself. Since the weights no longer
distinguish the partitions, the sum over $\sigma$ rebuilds the ordinary moment
$m_n$ from the Boolean cumulants. At long lags the $G$-cumulant and the
correlation function of order $n$ both reduce to $m_n\,P_1(\Delta)$, the
contribution of the histories in which the noise stays frozen over the whole
span at a value drawn from $p(\xi)$. All orders therefore decay in the same
way, as $P_1(\Delta)$, i.e.\ as $L(\Delta)\,\Delta^{-(\mu-2)}$ for $\mu>2$ and
as $(T/\Delta)^{\mu-2}$ for Eq.~\eqref{eq:manneville}, and differ only in the
amplitude $m_n$. The waiting-time law survives only through $P_1(\Delta)$,
i.e., asymptotically, through $\mu$, $L$ and $\tau$.
\end{inwords}

\noindent\emph{Proof outline.} By Eq.~\eqref{eq:lemma_D}, the difference
$|D_\sigma-P_1(\Delta)|$ is bounded by the two-epoch terms
$P_1(\ell_1)\Psi(\ell_p)$, $P_1(\ell_1)P_1(\ell_p)$ and $P_1(\ell_*)^2$. For a
regularly varying tail, Karamata's theorem makes $P_1$ regularly varying, with
index $-(\mu-2)$, and $\Psi(x)/P_1(x)=O(1/x)$. Since all the spans $\ell_k$
grow in proportion to $\Delta$, each two-epoch term is
$O\big(P_1(\Delta)^2\big)$, and dividing by $P_1(\Delta)$ gives
Eq.~\eqref{eq:crossover}. Inserted in Eq.~\eqref{eq:c_n}, together with
$\sum_\sigma\prod_Bb_{|B|}=m_n$, this gives $c_n=m_n+O\big(P_1(\Delta)\big)$;
Eq.~\eqref{eq:lemma_A}, inserted in Eq.~\eqref{eq:Cn_general}, gives the same
for $C_n$. The complete proof is given in Appendix~\ref{app:thm_crossover}.

\medskip
A numerical check at sixth order is reported in Section~\ref{sec:n6}; it
shows that the rate in Eq.~\eqref{eq:crossover} is attained. The theorem also covers the borderline case $\mu=2$ with a finite mean
waiting time, for which $P_1$ is slowly varying and decays more slowly than
any power of $\Delta$. For instance,
$\Psi(\theta)=1/[(1+\theta)\ln^3(e+\theta)]$ gives $P_1(\Delta)\sim(\ln\Delta)^{-2}/(2\tau)$:
the crossover still occurs, but at a logarithmic rate. For $\mu<2$ there is no
stationary regime, and the relevant comparison is with the age of the process;
this case is not considered here.

For power-law tails the long-lag form~\eqref{eq:universal} of the
$G$-cumulant is not, by itself, unexpected. Inserting the form
$C_n\simeq m_nP_1(\Delta)$ of Ref.~\cite{bblmCSF202} in the inversion~\eqref{eq:kappaG_def} gives it at once: since
$\overline\xi=0$, every product of two or more correlation functions contains
at least two single-epoch factors whose spans grow in proportion to $\Delta$,
and is therefore $O\big(P_1(\Delta)^2\big)$. The content of
Theorem~\ref{thm:crossover} lies elsewhere. First, it extends this behavior,
for both $C_n$ and the $G$-cumulants, to a general regularly varying tail,
including $\mu=2$, with an explicit rate. Second, and more importantly, it
concerns the individual renewal weights $D_\sigma$, which are properties of
the renewal process alone. They cannot be obtained from the long-lag form of
the $G$-cumulants, even by varying the jump law, because partitions with the
same block sizes, such as $\{12\}\{3456\}$ and $\{1234\}\{56\}$, share the
same coefficient $b_2b_4$, so that only their sum is fixed. The weights with
$p\ge2$ blocks distinguish the waiting-time laws: they are identically zero
for Poissonian renewal (Corollary~\ref{cor:poisson}), and asymptotically
equal to $P_1(\Delta)$ for power-law tails.

\paragraph{Beyond regular variation.} The proof only requires multi-epoch
coverings to be negligible compared with the single-epoch probability;
regular variation is one sufficient realization of this mechanism. The
bounds~\eqref{eq:lemma_A} and~\eqref{eq:lemma_D} hold for every waiting-time
law with bounded renewal density, and imply the crossover whenever the
two-epoch terms they contain are negligible compared with $P_1(\Delta)$
(Appendix~\ref{app:thm_crossover}).

\subsection{A numerical check of the rate\label{sec:n6}}

At $n=6$, with equal gaps $d_i=\Delta/5$, the interval partitions without
singleton blocks other than $1_6$ are $\{12\}\{3456\}$, $\{1234\}\{56\}$,
$\{123\}\{456\}$ and $\{12\}\{34\}\{56\}$. A two-block weight depends only on
the end points of its blocks,
\[
D_{\{B_1\}\{B_2\}}=A_{\{B_1\}\{B_2\}}-\Pr[E_{B_1}]\,\Pr[E_{B_2}],
\]
and is therefore the fourth-order quantity $P_1\rho_4$ of Eq.~\eqref{eq:rho4},
evaluated at the four times $t_{\min B_1}$, $t_{\max B_1}$, $t_{\min B_2}$ and
$t_{\max B_2}$; it follows from Eq.~\eqref{eq:A1234}. We computed the
two-block weights exactly for Pareto/Manneville waiting times
(Appendix~\ref{app:numerics}). Table~\ref{tab:n6} collects the results for
$\mu=2.5$. The normalized weights tend to one, and the deficit
$(1-r_\sigma)/P_1(\Delta)$ tends to a constant, as stated by
Theorem~\ref{thm:crossover}: the rate of Eq.~\eqref{eq:crossover} is
attained. For $\mu=3.5$ the deficits are much smaller, and
$(1-r_{\{12\}\{3456\}})/P_1(\Delta)=20.2$ at $\Delta=3000\,T$.

\begin{table}[t]
\centering
\caption{Normalized weights $r_\sigma=D_\sigma/P_1(\Delta)$ of the two-block
interval partitions without singleton blocks at sixth order, for stationary
Pareto/Manneville waiting times with $\mu=2.5$ and equal gaps $d_i=\Delta/5$
(exact calculation from Eq.~\eqref{eq:A1234}, Appendix~\ref{app:numerics}).
By symmetry, $r_{\{1234\}\{56\}}=r_{\{12\}\{3456\}}$. The last column stays
bounded, as required by Theorem~\ref{thm:crossover}.}
\label{tab:n6}
\smallskip
\begin{tabular}{rcccc}
\hline
$\Delta/T$ & $P_1(\Delta)$ & $r_{\{12\}\{3456\}}$ & $r_{\{123\}\{456\}}$ &
$\dfrac{1-r_{\{12\}\{3456\}}}{P_1(\Delta)}$\\[1ex]
\hline
$30$   & $0.180$  & $0.541$ & $0.590$ & $2.56$\\
$100$  & $0.0995$ & $0.730$ & $0.763$ & $2.71$\\
$300$  & $0.0576$ & $0.840$ & $0.860$ & $2.77$\\
$1000$ & $0.0316$ & $0.912$ & $0.923$ & $2.80$\\
$3000$ & $0.0183$ & $0.949$ & $0.955$ & $2.81$\\
\hline
\end{tabular}
\end{table}
\section{The long-lag memory kernel and master equation\label{sec:ht_kernel}}

\begin{summary}
At long lags the $G$-cumulants of the noise can be replaced by $m_nP_1$, and
the memory kernel can then be summed at all orders. It becomes the
probability that the whole memory interval lies in one epoch, times a
second-order kernel in which the system evolves under the frozen flow at the
value held by the noise, averaged with the weight $\xi^2p(\xi)$. The
Poissonian kernel has the same structure, but with the Boolean L\'evy measure
in place of $\xi^2p(\xi)$. The corresponding master equation is written in
closed form, and the conditions for its validity are discussed.
\end{summary}

Suppose that all the lags that
contribute to the kernel are large compared with the waiting-time scale $T$.
Theorem~\ref{thm:crossover} then suggests replacing the $n$-th $G$-cumulant with $m_nP_1(\Delta)$, where now
$\Delta=t-u$ is the kernel lag. In the Schr\"odinger picture the $n$-th term
of the kernel then reads
\begin{equation}
\mathcal G_n(\Delta)\simeq\epsilon^n m_n\,P_1(\Delta)
\int_{\substack{d_i>0\\\sum d_i=\Delta}}
\mathcal L_I\,e^{\mathcal L_ad_1}\mathcal L_I\cdots e^{\mathcal L_ad_{n-1}}\mathcal L_I
\prod_i\mathrm dd_i .
\label{eq:kernel_tail}
\end{equation}
This should be compared with the exact Poissonian chain, in which $m_n$ is
replaced by $b_n$ and $P_1(\Delta)$ by $e^{-\Delta/\tau}$, and which sums to
the frozen-flow form~\eqref{eq:kernel_poisson_rho}, weighted by the Boolean
L\'evy measure. The power-law chain~\eqref{eq:kernel_tail} sums in the same
way, but with a different measure.

\begin{result}[Long-lag frozen-flow representation of the kernel]
\label{prop:spectral}
Assume the factorized joint law~\eqref{eq:joint_law_factorized}. For
power-law waiting times with finite mean, in the regime in which the
replacement~\eqref{eq:kernel_tail} holds,
\begin{equation}
\mathcal G(\Delta)\simeq\epsilon^2\,P_1(\Delta)\,
\mathcal L_I\Big[\int_{-\infty}^{\infty}\xi^2e^{\mathcal L_\xi\Delta}p(\xi)\,\mathrm d\xi\Big]\mathcal L_I .
\label{eq:kernel_powerlaw_p}
\end{equation}
\end{result}

\begin{inwords}
At long lags the kernel is dominated by the histories in which the noise
holds one of its actual values over the whole memory interval, and it becomes
an average over the corresponding frozen flows. It is the probability
$P_1(\Delta)$ that the whole memory interval lies in one epoch, times a
second-order kernel in which the system evolves under the frozen flow at the
value $\xi$, averaged with the weight $\xi^2p(\xi)$. The Poissonian
kernel~\eqref{eq:kernel_poisson_rho} has the same structure, and is exact at
all lags, but its measure is the Boolean L\'evy density $\varrho$. Each frozen
propagator contains all orders in $\epsilon$.
\end{inwords}

\noindent\emph{Derivation.} The chain~\eqref{eq:kernel_tail} with $n=k+2$
couplings is the $k$-th term of the Dyson expansion~\eqref{eq:dyson},
sandwiched between two factors $\epsilon\mathcal L_I$ and averaged over the
measure $\omega(\xi)=\xi^2p(\xi)$, whose moments are
$\int_{-\infty}^{\infty}\xi^k\omega(\xi)\,\mathrm d\xi=m_{k+2}$, with the factor
$P_1(\Delta)$ in place of $e^{-\Delta/\tau}$. Summing over $k$ gives
Eq.~\eqref{eq:kernel_powerlaw_p}. $\square$

\medskip
Used as the kernel of Eq.~\eqref{eq:ME}, Eq.~\eqref{eq:kernel_powerlaw_p}
gives for Eq.~\eqref{eq:SDE}, under stationary forcing, the moment-weighted
master equation
\begin{equation}
\partial_tP(x\pv t)=\partial_x\big[C(x)P(x\pv t)\big]
+\epsilon^2\int_{t_0}^t\mathrm du\,P_1(t-u)\int_{-\infty}^{\infty}\mathrm d\xi\,\xi^2p(\xi)\,
\partial_x\Big\{I(x)\,e^{\mathcal L_\xi(t-u)}\,\partial_x\big[I(x)P(x\pv u)\big]\Big\},
\label{eq:ME_mom}
\end{equation}
with $\mathcal L_\xi\rho=\partial_x[(C-\epsilon\xi I)\rho]$. It contains all
orders in $\epsilon$ and is correct at long lags; its accuracy as a complete
kernel is tested in Section~\ref{sec:limi}.

The Poissonian kernel~\eqref{eq:kernel_poisson_rho} and the power-law
long-lag kernel~\eqref{eq:kernel_powerlaw_p} thus differ only in the measure
that weights the frozen flows. We call kernels of this form extended
telegraph kernels: for dichotomous noise and Poissonian renewal,
Eq.~\eqref{eq:kernel_poisson_rho} reduces to the second-order telegraph kernel
(Section~\ref{sec:poisson_kernel}). In the Poissonian kernel the measure is the
Boolean L\'evy density $\varrho(x)$, which is not the law of the values held by
the noise. In the power-law kernel it is
$\xi^2p(\xi)$: the noise holds its actual value $\xi$ over the whole memory
interval, and the factor $\xi^2$ comes from the two couplings
$\epsilon\xi\mathcal L_I$ at the ends of the interval. The two measures have
the same mass, $b_2=\overline{\xi^2}$, and the same large-$|x|$ behavior,
since $\varrho(x)\simeq x^2p(x)$~\cite{bmBoolean}; they differ in the bulk.
Their moments are $b_{n+2}$ and $m_{n+2}$, so the crossover $c_n\to m_n$ of
Theorem~\ref{thm:crossover} is, at the level of the
kernel, a crossover of the frozen-flow measure from $\varrho(x)$ to
$x^2p(x)$.

The dichotomous case makes the difference concrete. For power-law tails the
weight is $\xi^2p=\frac{a^2}{2}(\delta_a+\delta_{-a})$: instead of the free
propagator of the telegraph kernel, the carrier evolves under the frozen flows
at $\pm a$, which contain all orders in $\epsilon$. This is the
dynamical content of Corollary~\ref{cor:closure}.

In Laplace space the product with $P_1(\Delta)$ in
Eq.~\eqref{eq:kernel_powerlaw_p} is a convolution, not the shift
$s\to s+1/\tau$ produced by $e^{-\Delta/\tau}$ in the Poissonian kernel, and
no resolvent form follows. Since the frozen propagator is an exponential,
however, the transform is still explicit, as a function of the frozen
generator:
\begin{equation}
\hat{\mathcal G}(s)\simeq\epsilon^2\,
\mathcal L_I\Big[\int_{-\infty}^{\infty}\xi^2\,\hat P_1(s-\mathcal L_\xi)\,p(\xi)\,\mathrm d\xi\Big]\mathcal L_I,
\qquad
\hat P_1(s-\mathcal L_\xi):=\int_0^\infty P_1(\Delta)\,e^{(\mathcal L_\xi-s)\Delta}\,\mathrm d\Delta .
\label{eq:kernel_powerlaw_laplace}
\end{equation}
For Poissonian renewal, $\hat P_1(z)=\tau/(1+z\tau)$
turns $\hat P_1(s-\mathcal L_x)$ into $(s+1/\tau-\mathcal L_x)^{-1}$, the
resolvent of Ref.~\cite{bmBoolean}. For power-law tails $\hat P_1(z)$ is singular
at $z=0$, and each frozen generator moves this singularity to its non-zero
eigenvalues, the slowest of which is $-\lambda(\xi)$ (the zero eigenvalue, of
the frozen stationary density, does not contribute, since
$\int\mathcal L_I\varphi\,\mathrm dx=0$ for every $\varphi$). The singularities of the long-lag
kernel are therefore branch points at the frozen rates, inherited from
$\hat P_1$, whereas those of the Poissonian resolvent are poles, shifted by
$-1/\tau$. In the first-moment sector of the LIMI/CAM model,
Eq.~\eqref{eq:kernel_powerlaw_laplace} reduces to the kernel
$\tilde K_{\rm mom}$ used in Section~\ref{sec:limi}. The atom surrogates of Ref.~\cite{bmBoolean}
remain admissible. They replace $p$ by its $N$-point Gauss rule $p_N$ in
Eq.~\eqref{eq:kernel_powerlaw_p}, which preserves the moments $m_n$, and hence
the chain coefficients, up to $n=2N-1$ (Remark~\ref{rem:atoms}). For
Poissonian renewal, instead, they replace $\varrho$ by its $(N-1)$-point Gauss
rule~\cite{bmBoolean}.

\paragraph{Validity of the replacement.} Equation~\eqref{eq:kernel_tail}
requires that the chain integral not be dominated by configurations with gaps
of order $T$, where $c_n$ is not yet close to $m_n$. In the free case
($\mathcal L_a=0$, e.g.\ the L\'evy walk) the chain integrand is uniform over
the simplex of intermediate times, and the replacement amounts to
$\langle c_n\rangle_{\rm simplex}\to m_n$. At fourth order this is
$\langle\rho_4\rangle_{\rm simplex}\to1$, which we verified with the exact
renewal calculation of Appendix~\ref{app:numerics}: for Manneville waiting
times with $\mu=3.5$,
$\langle\rho_4\rangle_{\rm simplex}=0.755,\,0.926,\,0.980,\,0.995$ for
$\Delta/T=25,\,100,\,400,\,1600$. The deficit decreases approximately as
$T/\Delta$, more slowly than $P_1(\Delta)\propto(T/\Delta)^{3/2}$: on the
simplex it is dominated by configurations in which one outer gap is of order
$T$. For a dissipative dynamics the relevant
lags are of the order of the relaxation time of the frozen flows, and the
replacement requires this time to be long compared with $T$. This is tested on
an exactly solvable model in Section~\ref{sec:limi}.
\section{A test: the stationary mean of the LIMI/CAM model}
\label{sec:limi}

\begin{summary}
The LIMI/CAM model provides an exact dynamical test of the kernel theory.
Its stationary mean can be obtained exactly for arbitrary factorized jump and waiting-time laws. The
model shows that slow resolved dynamics does not guarantee a second-order
closure: for power-law waiting times with $2<\mu<3$, the relative error
remains finite in the slow-dynamics limit.
\end{summary}

Let the LIMI/CAM model~\eqref{eq:limi} be driven by stationary step-renewal
noise, in the stable regime. On the first-moment sector the adjoint generators act as $\mathcal A_ax=-\gamma x$ and
$\mathcal A_If=(1+\beta x)f'$. Every chain in the kernel then reduces to the
scalar factor $\beta^{n-1}e^{-\gamma\Delta}(1+\beta x)$. The dissipative drift
enters only through an overall factor $e^{-\gamma\Delta}$, and the chain
integrand is uniform over the simplex of intermediate times. The mean
$m(t)=\langle x\rangle$ therefore obeys
\begin{align}
\dot m&=-\gamma m+\epsilon^2\beta\int_0^t\mathrm d\Delta\,K(\Delta)\,
\big[1+\beta\,m(t-\Delta)\big],
\nonumber\\
K(\Delta)&=e^{-\gamma\Delta}\sum_{n\ge2}(\epsilon\beta)^{n-2}
\int_{\rm simplex}\doubleangle{\xi(t-\Delta)\cdots\xi(t)}^{(G)} ,
\label{eq:limi_mean_eq}
\end{align}
and its stationary value is
$m_{\rm st}=\epsilon^2\beta\tilde K/(\gamma-\epsilon^2\beta^2\tilde K)$, with
$\tilde K=\int_0^\infty K$.

\paragraph{Exact result.} The kernel~\eqref{eq:limi_mean_eq} sums to an exact
expression. Integrating Eq.~\eqref{eq:limi} along a trajectory,
\[
m_{\rm st}=\epsilon\int_0^\infty\mathrm dv\,e^{-\gamma v}
\Big\langle\xi(-v)\,e^{\epsilon\beta\int_{-v}^0\xi}\Big\rangle ,
\]
and since $\xi(-v)\,e^{\epsilon\beta\int_{-v}^0\xi}
=(\epsilon\beta)^{-1}\partial_v\,e^{\epsilon\beta\int_{-v}^0\xi}$, an
integration by parts gives
\begin{equation}
m_{\rm st}=\frac{\gamma\,\hat\Phi(\gamma)-1}{\beta},
\qquad
\Phi(v):=\Big\langle e^{\epsilon\beta\int_0^v\xi(t)\,\mathrm dt}\Big\rangle .
\label{eq:limi_exact}
\end{equation}
For a stationary renewal process, $\hat\Phi$ follows from the renewal
decomposition at the first renewal,
\begin{equation}
\hat\Phi(s)=\big\langle\hat P_1(s-\epsilon\beta\xi)\big\rangle_\xi
+\big\langle\hat f_e(s-\epsilon\beta\xi)\big\rangle_\xi\,
\frac{\big\langle\hat\Psi(s-\epsilon\beta\xi)\big\rangle_\xi}
{1-\big\langle\hat\psi(s-\epsilon\beta\xi)\big\rangle_\xi},
\label{eq:Phi_hat}
\end{equation}
where hats denote Laplace transforms. The first term is the contribution of
trajectories with no renewal in $[0,v]$. In the second, the first renewal
occurs after a forward-recurrence time and the process then restarts from a
renewal. At $s=\gamma$ all the transforms are evaluated at
$s-\epsilon\beta\xi=\lambda(\xi)>0$, where they converge because of the
stability condition. Equations~\eqref{eq:limi_exact} and~\eqref{eq:Phi_hat}
are exact for any waiting-time law and any jump law under
amplitude--duration factorization. For a non-factorized joint law, the
separate averages in Eq.~\eqref{eq:Phi_hat} must be replaced by joint
amplitude--duration transforms.

\paragraph{Kernel predictions.} In this sector,
Eqs.~\eqref{eq:kernel_poisson_rho} and~\eqref{eq:kernel_powerlaw_p}, with
$e^{-\Delta/\tau}$ replaced by $P_1(\Delta)$ in the former, give
\begin{equation}
\tilde K_{\rm Bool}=\int_{-\infty}^{\infty}\varrho(x)\hat P_1\big(\lambda(x)\big)\,\mathrm dx,
\qquad
\tilde K_{\rm mom}=\big\langle\xi^2\,\hat P_1\big(\lambda(\xi)\big)\big\rangle_\xi ,
\label{eq:limi_kernels}
\end{equation}
where for Poissonian renewal $\hat P_1(z)=\tau/(1+z\tau)$. The second-order
truncation gives $\tilde K_2=b_2\,\hat P_1(\gamma)$, which for the symmetric
Bernoulli law coincides with $\tilde K_{\rm Bool}$. We denote by
$m_{\rm Bool}$, $m_{\rm mom}$ and $m_2$ the corresponding stationary means.

\begin{figure}[t]
\centering
\includegraphics[width=\textwidth]{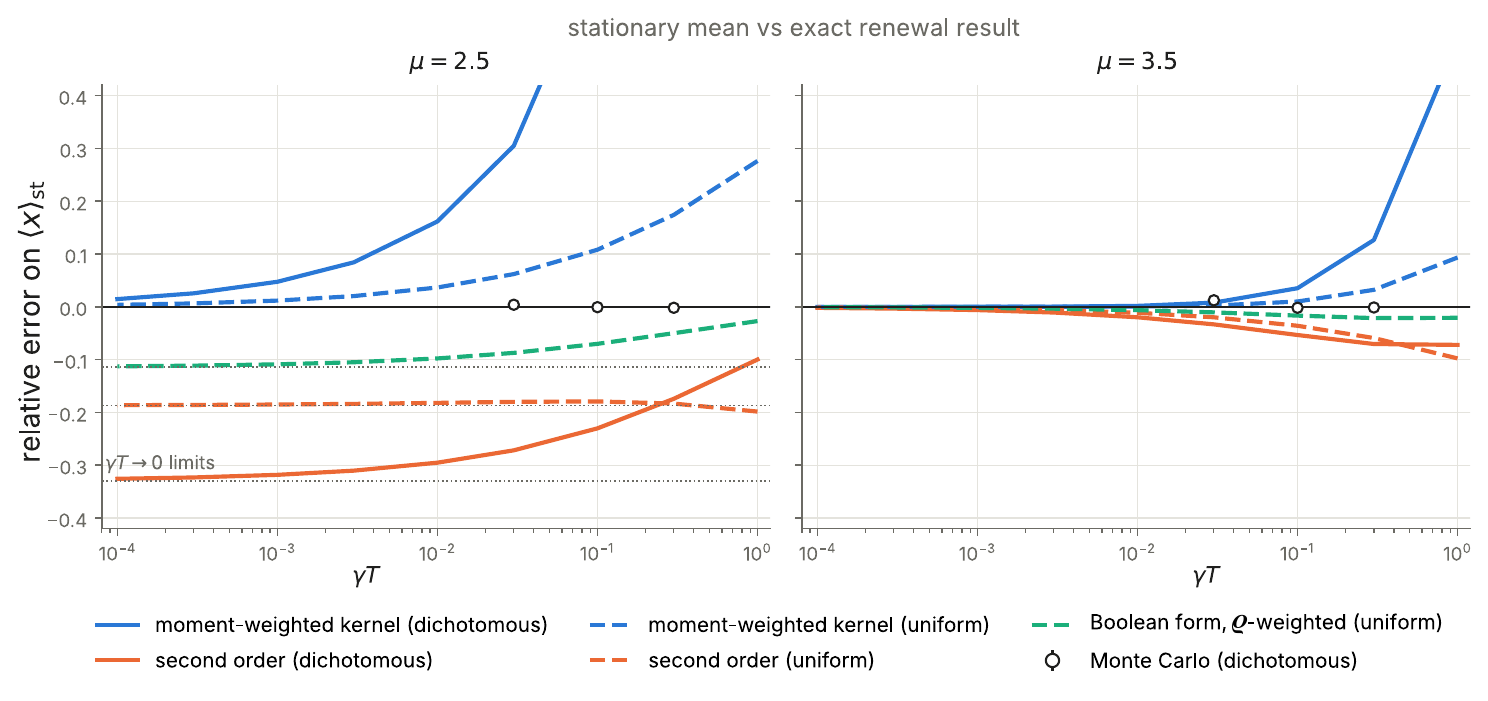}
\caption{Stationary mean of the LIMI/CAM model~\eqref{eq:limi} driven by
step-renewal noise with Manneville waiting times ($T=1$), with $\mu=2.5$
(left) and $\mu=3.5$ (right), for $\epsilon\beta\max|\xi|=0.8\,\gamma$ and
$\beta=1$: relative error of the moment-weighted kernel of
Result~\ref{prop:spectral} (blue), of the second-order truncation (orange)
and of the $\varrho$-weighted Boolean form (green; for dichotomous jumps it
coincides with the second-order truncation), with respect to the exact
result~\eqref{eq:limi_exact}. Solid lines: dichotomous jumps; dashed lines:
uniform jumps. Dotted lines: the limits~\eqref{eq:limi_limits}. Symbols:
Monte Carlo simulation of Eq.~\eqref{eq:limi} (dichotomous jumps,
$6\times10^5$--$10^6$ trajectories), relative to the exact result.}
\label{fig:limi}
\end{figure}

\paragraph{Results.} For exponential waiting times,
Eq.~\eqref{eq:limi_exact} and the Boolean form~\eqref{eq:limi_kernels} agree to
seven significant figures for both dichotomous and uniform jumps, a dynamical
check of Eq.~\eqref{eq:kernel_poisson_rho}. For Manneville waiting times,
Fig.~\ref{fig:limi} compares the three approximations with the exact result,
at fixed coupling ratio $\epsilon\beta\max|\xi|=0.8\,\gamma$, as the dynamics is
slowed down ($\gamma T\to0$). A direct Monte Carlo simulation of
Eq.~\eqref{eq:limi} with dichotomous jumps, with exact integration inside each
epoch and up to $10^6$ trajectories, agrees with Eq.~\eqref{eq:limi_exact}
within $1.5$ standard deviations at all the points tested.
\begin{itemize}
\item The moment-weighted kernel~\eqref{eq:kernel_powerlaw_p} becomes exact as
$\gamma T\to0$, with a relative error decaying as $(\gamma T)^{\mu-2}$ for
$\mu=2.5$. This is the regime in which the lags that matter, of order
$1/\lambda(\xi)$, are much longer than $T$.
\item For $2<\mu<3$ the second-order truncation and the Boolean form do
\emph{not} become exact. In the limit $\gamma T\to0$ at fixed coupling ratio,
$\hat P_1(z)\simeq\Gamma(3-\mu)\,T^{\mu-2}z^{\mu-3}$. The term
$\epsilon^2\beta^2\tilde K$, of relative order $(\gamma T)^{\mu-2}$, becomes
negligible against $\gamma$, and
\begin{equation}
\frac{m_2}{m_{\rm st}}\to
\frac{b_2}{\big\langle\xi^2\,[\lambda(\xi)/\gamma]^{\mu-3}\big\rangle_\xi},
\qquad
\frac{m_{\rm Bool}}{m_{\rm st}}\to
\frac{\int_{-\infty}^{\infty}\varrho(x)[\lambda(x)/\gamma]^{\mu-3}\,\mathrm dx}
{\big\langle\xi^2\,[\lambda(\xi)/\gamma]^{\mu-3}\big\rangle_\xi},
\label{eq:limi_limits}
\end{equation}
i.e.\ relative errors of $-33\%$ (dichotomous), $-19\%$ (uniform, second order)
and $-11\%$ (uniform, Boolean form) for $\mu=2.5$ (dotted lines in
Fig.~\ref{fig:limi}).
\item For $\mu>3$ all the approximations converge as $\gamma T\to0$.
\end{itemize}
The dividing line $\mu=3$ is where the correlation time of the noise,
$\int_0^\infty P_1(\Delta)\,\mathrm d\Delta$, becomes finite. For $\mu>3$ slow
dynamics sees an effectively short-correlated noise, and the Poissonian
intuition holds: the higher orders of the kernel become negligible. For
$2<\mu<3$ the correlation time is infinite. The fourth- and higher-order
contributions then remain of the same order as the second-order one however
slow the dynamics, and the stationary state is controlled by the
moment-weighted kernel of Result~\ref{prop:spectral}. A separation of time
scales, $\gamma T\to0$, does not rescue a second-order truncation in this
regime. Both jump laws used here lie below the amplitude
$\xi_{\rm c}=\gamma/(\epsilon\beta)$ at which frozen flows become unstable.

\section{Discussion\label{sec:discussion}}

\begin{summary}
The results can be read directly from the master equations derived here.
Several of them confirm, for the reduced dynamics, known features of
single-big-jump physics; the exact decomposition, the all-order crossover of
the kernel coefficients, the frozen-flow master equation and the failure of
second-order closures for $2<\mu<3$ are new.
\end{summary}

At the second order, the law governing waiting times enters the master equation
(Eq.~\eqref{eq:ME2}) solely through the single-epoch probability $P_1$; a second-order closure is exact only in the case of the telegraph process
(Eq.~\eqref{eq:ME_telegraph}). For waiting times that follow a power law, higher-order
terms do not become negligible as the lag increases: for large lags, all coefficients
decay in the same manner as $P_1(\Delta)$, differing only by an amplitude $m_n$. The
kernel does not converge toward a second-order Gaussian form, but rather toward a kernel
associated with a randomly frozen driving force (Eq.~\eqref{eq:kernel_powerlaw_p}). The
master equation (Eq.~\eqref{eq:ME_mom}) provides a physical expression of this: over a
long memory interval, the system---with probability $P_1$---has been subjected to a single
value of the driving force, drawn from the distribution $p(\xi)$, and has evolved
accordingly, following the corresponding deterministic flow. The form of the kernel that
determines the reduced density depends on the ratio between the relaxation time of the
dynamics and the characteristic timescale $T$ of the waiting times. For lags that are
short compared to $T$, the coefficients reduce to Boolean cumulants, regardless of the
specific waiting-time law (Eq.~\eqref{eq:short_lag}); conversely, slow dynamics involve an
integration of the kernel over lags that are long compared to $T$, and are thus governed
by the asymptotic form applicable to large lags. In the case $2<\mu<3$, the correlation
time $\int_0^\infty P_1$ is infinite; consequently, the higher-order terms remain of the
same order of magnitude as the second-order term---regardless of how slow the dynamics
are---which is why the separation of time scales is insufficient to justify a second-order
closure. For $\mu>3$, on the other hand, the correlation time is finite, and in the limit
of slow dynamics, the second-order closure becomes asymptotically exact.

\paragraph{Relation to known results.} Some of these features independently confirm
known physical mechanisms. The fact that rare events are dominated by a single,
exceptional contribution corresponds to the single-big-jump principle characteristic of
subexponential distributions~\cite{foss2013,embrechts1997}---a principle used to model rare
events in Lévy walks, cold-atom systems, quenched disordered media, and correlated
walks~\cite{vbbPRE100,wvbbPRR1,hbEPJB94}. The mechanism by which individual long-duration
events---rather than the prevailing statistical behavior---determine high-order moments lies
at the heart of infinite densities and strongly anomalous
diffusion~\cite{rdhbPRE90,vbbPRE100}; the crossover of the kernel coefficients toward the
moments represents the equivalent of this within reduced dynamics. The asymptotic form
$C_n\simeq m_nP_1(\Delta)$ for correlation functions, in the case of waiting times that
follow a power law, was derived in~\cite{bblmCSF202}; the Boolean structure for Poissonian
renewal processes is described in~\cite{bmBoolean}, while the exact second-order closure
for dichotomous Markovian noise is a classic
result~\cite{slPhysicaA91,Horsthemke1984,bIJMPB20}. The findings that, to the best of our
knowledge, are new include: the exact decomposition of the $G$-cumulants, for an arbitrary
waiting-time distribution, into Boolean cumulants of the jump distribution and renewal
weights; the proof of the crossover at every order for the individual renewal weights,
with an explicit rate; the expression of the long-lag kernel as a superposition of
frozen-flow kernels, together with the corresponding master equation; the
inadequacy of second-order closures in the range $2<\mu<3$, regardless of the separation
between time scales; and, finally, the uniqueness of the telegraph process as an exact
second-order closure.

\section{Conclusions\label{sec:conclusions}}

For a system driven by renewal step noise with mutually independent amplitudes and
durations, the coefficients of the memory kernel in the master equation can
be decomposed exactly into Boolean cumulants of the jump distribution and weights that
depend solely on the renewal process. At second order, the kernel depends on the waiting
times only through the single-epoch probability $P_1$; the telegraph process is the only
renewal step noise for which the kernel closes exactly at second order. In the case of waiting times that follow a power law with a finite
mean, the coefficients---at all orders and with relative corrections of the order of
$P_1$---transition from Boolean cumulants to the ordinary moments of the jump distribution
for long time intervals. Consequently, the kernel for long intervals is given by the
probability $P_1$ (the probability that the memory interval falls within a single epoch),
multiplied by the kernel of the frozen flow at the value assumed by the noise, and
averaged with the weight $\xi^2p(\xi)$; in this way, the master equation takes
the closed form~\eqref{eq:ME_mom}. In the LIMI/CAM model, for $2<\mu<3$, a second-order
closure leads to a finite error in the stationary mean, regardless of how slow the
dynamics are; in contrast, the master equation~\eqref{eq:ME_mom} becomes exact in this
limit.
Consequently, in the case of intermittent forcing whose durations follow a power
law, the separation of time scales does not justify a Gaussian description---nor a
second-order description---of the driven variable: for slow dynamics, the appropriate
reduced equation is constructed on the frozen flows, which are themselves weighted
according to the law governing the jumps.

The single-epoch mechanism entails further consequences, which will be
discussed separately and upon which the results presented above do not depend. Although
stretched-exponential waiting times exhibit a finite correlation time, long-duration
epochs remain significantly more probable than they would be under a purely exponential
law; this raises the question of whether they belong to the same class. In the presence of
exponential tails, the Boolean structure can re-emerge over extended time intervals. The
relaxation of the driven system and its stationary density---including its support,
singularities, and tails---are determined by the frozen flows and the extreme values of
the law governing the jumps; the same applies to the statistics of strong events. Finally,
correlated amplitudes and durations---such as those encountered, for instance, in Sisyphus
cooling~\cite{vbbPRE100}---as well as vector forcing, represent natural extensions of the
theoretical framework presented here.

\section*{CRediT authorship contribution statement}
\textbf{M.~Bianucci:} Conceptualization, Methodology, Formal analysis,
Software, Writing -- original draft. \textbf{R.~Mannella:} Validation,
Writing -- review \& editing.

\section*{Declaration of competing interest}
The authors declare that they have no known competing financial interests or
personal relationships that could have appeared to influence the work
reported in this paper.

\section*{Declaration of generative AI and AI-assisted technologies in the writing process}
During the preparation of this work the authors used Claude (Anthropic) in
order to assist with language editing and the organization of the text, and
to check derivations and numerical calculations. After using this tool, the
authors reviewed and edited the content as needed and take full
responsibility for the content of the publication.

\section*{Data availability}
The codes used for all the calculations and figures of this work are
available from the corresponding author upon reasonable request.

\appendix
\section{Proofs\label{app:proofs}}

\subsection{Two-epoch bounds\label{app:two_plateau}}

\begin{lemma}[Two-epoch bound]
\label{lem:two_plateau}
Consider a stationary renewal process with $\tau<\infty$ and bounded renewal
density, $\|R\|_\infty:=\sup_{t>0}R(t)<\infty$. For every
$\sigma\in\mathcal I(n)$ with $p\ge2$ blocks, the bounds~\eqref{eq:lemma_A}
and~\eqref{eq:lemma_D} hold.
\end{lemma}

\begin{proof}
The lower bound in Eq.~\eqref{eq:lemma_A} holds because a single epoch
 over $[t_1,t_n]$ satisfies all the constraints. The difference
$A_\sigma-P_1(\Delta)$ is the probability that no renewal occurs inside any
block, but at least one occurs in $(t_1,t_n)$. On this event the first renewal
after $t_1$, $u_1$, lies in $(t_1+\ell_1,t_{\min B_p})$, and so does the last
renewal before $t_n$, $u\ge u_1$, whose waiting time must exceed
$t_n-u\ge\ell_p$. Dropping the constraints on the intermediate blocks,
\[
A_\sigma-P_1(\Delta)\le\int_{t_1+\ell_1}^{t_{\min B_p}}\mathrm du_1\,
f_e(u_1-t_1)\int_{u_1}^{t_{\min B_p}}\mathrm du\,\tilde R(u-u_1)\,
\Psi(t_n-u).
\]
Since $\Psi$ is non-increasing and $t_n-u_1\ge\ell_p$, the inner integral is at
most $\Psi(\ell_p)+\|R\|_\infty\int_{\ell_p}^\infty\Psi
=\Psi(\ell_p)+\tau\|R\|_\infty P_1(\ell_p)$, and the integral of $f_e$ over
$u_1\ge t_1+\ell_1$ is at most $P_1(\ell_1)$. This proves
Eq.~\eqref{eq:lemma_A}. In Eq.~\eqref{eq:kappaG_general} the coarsest
partition $1_n$ contributes $A_\sigma$. The other $2^{p-1}-1$ coarsenings of
$\sigma$ contribute products of at least two factors, and each factor
satisfies $A_{\sigma|_Z}\le\Pr[E_B]=P_1(\ell_B)\le P_1(\ell_*)$ for any block
$B\subseteq Z$ of $\sigma$. Each such product is therefore at most
$P_1(\ell_*)^2$, which together with Eq.~\eqref{eq:lemma_A} gives
Eq.~\eqref{eq:lemma_D}.
\end{proof}

The renewal density is bounded whenever
$\sum_{k\ge0}\sup_{[k,k+1]}\psi<\infty$, in particular when $\psi$ is
directly Riemann integrable, or bounded and eventually non-increasing. Indeed,
$R(t)=\int_{[0,t]}\psi(t-s)\,U(\mathrm ds)$, with
$U(\mathrm ds)=\delta(s)\,\mathrm ds+R(s)\,\mathrm ds$ the renewal measure, and
$U([x,x+1])\le U([0,1])$, because the renewals in $[x,x+1]$ are those of a
process started at the first of them. Hence
$\|R\|_\infty\le U([0,1])\sum_{k\ge0}\sup_{[k,k+1]}\psi$.

\subsection{Proof of Theorem~\ref{thm:crossover}\label{app:thm_crossover}}

\begin{proof}
For $\mu>2$, Karamata's theorem~\cite{bgtRV} gives
$\int_x^\infty\Psi\sim x^{-(\mu-2)}L(x)/(\mu-2)$. Hence
$P_1(x)\sim x^{-(\mu-2)}L(x)/[(\mu-2)\tau]$ is regularly varying with index
$-(\mu-2)$, and $\Psi(x)/P_1(x)\sim(\mu-2)\tau/x\to0$. For $\mu=2$ and
$\tau<\infty$, $P_1$ is slowly varying and again
$\Psi(x)/P_1(x)=o(1/x)$~\cite{bgtRV}. Since all the blocks have size $\ge2$, all the spans grow linearly,
$\ell_k=\alpha_k\Delta$, with fixed $\alpha_k>0$ (sums of gap fractions), so
that $P_1(\alpha\Delta)/P_1(\Delta)\to\alpha^{-(\mu-2)}$ for each of them. The
renewal density is bounded, as shown in Appendix~\ref{app:two_plateau}. Dividing Eq.~\eqref{eq:lemma_D} by
$P_1(\Delta)$,
\[
\Big|\frac{D_\sigma}{P_1(\Delta)}-1\Big|\le K_\sigma\,P_1(\Delta)\,[1+o(1)],
\qquad
K_\sigma:=\tau\|R\|_\infty\,(\alpha_1\alpha_p)^{-(\mu-2)}
+\big(2^{p-1}-1\big)\,\alpha_*^{-2(\mu-2)},
\]
with $\alpha_*=\min_k\alpha_k$, which is Eq.~\eqref{eq:crossover}.

The statement on $c_n$ follows from Eq.~\eqref{eq:c_n} and
$\sum_{\sigma}\prod_Bb_{|B|}=m_n$. For $C_n$, Eq.~\eqref{eq:lemma_A} gives
$A_\sigma=P_1(\Delta)[1+O(P_1(\Delta))]$ for every singleton-free $\sigma$,
and the result follows from Eq.~\eqref{eq:Cn_general}.
\end{proof}

The same argument shows that the bounds~\eqref{eq:lemma_A}
and~\eqref{eq:lemma_D} imply the crossover for every waiting-time law with
bounded renewal density for which the two-epoch probabilities are
negligible compared with the single-epoch one, i.e.\ whenever
\[
\frac{P_1(\ell_1)\big[\Psi(\ell_p)+P_1(\ell_p)\big]}{P_1(\Delta)}\to0
\qquad\text{and}\qquad
\frac{P_1(\ell_*)^2}{P_1(\Delta)}\to0 .
\]
For regularly varying tails both ratios are $O(P_1(\Delta))$, which gives the
rate of Theorem~\ref{thm:crossover}.

\section{Numerical methods\label{app:numerics}}

All the calculations use statistically independent jump amplitudes and
waiting times, as in the factorized model studied in the main text.

\paragraph{Renewal density.} On the grid $t_k=kh$ the renewal equation
$R=\psi+\psi*R$ is discretized with the trapezoidal rule,
\begin{equation}
R_k=\psi_k+h\Big[\sum_{j=1}^{k-1}\psi_{k-j}R_j
+\tfrac12\big(\psi_kR_0+\psi_0R_k\big)\Big],\qquad R_0=\psi_0 ,
\label{eq:app_volterra}
\end{equation}
with $\psi_k=\psi(t_k)$, and solved for $R_k$ step by step. To avoid the
$O(N^2)$ cost of the convolution sums, the grid is split recursively in two
halves; once the first half is solved, its contribution to the sums of the
second half is computed in a single FFT convolution. The total cost is
$O(N\ln^2N)$. Two solutions, with steps $h$ and $h/2$, are combined by
Richardson extrapolation, $R=(4R_{h/2}-R_h)/3$, which removes the $O(h^2)$
error of the trapezoidal rule. With $h=0.01\,T$ and $t\le2200\,T$ the scheme
reproduces $R\equiv1/\tau$ for Poissonian input to $4\times10^{-7}$.
\paragraph{Fourth-order weights.} In Eq.~\eqref{eq:A1234} the outer
integral over $u\in(t_2,t_3)$ is evaluated with a $1500$-point
Gauss--Legendre rule. The inner integral is split as
$\tilde R=\delta+1/\tau+(R-1/\tau)$. The $\delta$ part gives $\Psi(t_4-u)$,
the constant part gives $P_1(t_4-t_3)-P_1(t_4-u)$ analytically, and only the
small remainder $R-1/\tau$ is integrated numerically on the grid, by the
trapezoidal rule. On Poissonian input the resulting $\rho_4$ vanishes to
$10^{-6}$. This exact renewal calculation gives the simplex
averages of Section~\ref{sec:ht_kernel} and the two-block weights of
Table~\ref{tab:n6}.

\paragraph{Stationary mean of the LIMI/CAM model.} The Laplace transforms in
Eq.~\eqref{eq:Phi_hat} and in the kernels~\eqref{eq:limi_kernels} were
computed by adaptive quadrature on logarithmically spaced subintervals
(relative tolerance $10^{-11}$), to resolve the algebraic tails of $\Psi$ and
$P_1$. The Monte Carlo simulation of
Eq.~\eqref{eq:limi} integrates the frozen flow exactly inside each epoch,
starts from stationary noise and $x(t_0)=0$, and its convergence to
stationarity was checked by sampling at increasing times.

 \newcommand{\noop}[1]{}

\end{document}